\documentclass[article]{JHEP3}

\usepackage{amsmath,amssymb}
\usepackage[dvips]{graphics}
\usepackage{epsfig}
\usepackage{psfrag}

\newcommand{\re}{{\mathbb{R}}{\mathrm{e}}}
\newcommand{\im}{{\mathbb{I}}{\mathrm{m}}}
\newcommand{\be}{\begin{equation}}
\newcommand{\ee}{\end{equation}}
\newcommand{\bea}{\begin{eqnarray}}
\newcommand{\eea}{\end{eqnarray}}
\newcommand{\bean}{\begin{eqnarray*}}
\newcommand{\eean}{\end{eqnarray*}}

\def\beq{\begin{equation}}

\def\eeq{\end{equation}}

\relax

\preprint{{\small \texttt{hep-th/0605128}}}

\title{Perturbative Calculation of Quasinormal Modes of $d$--Dimensional
Black Holes}
\author{Fu-Wen Shu$^{a,c}$ and You-Gen
Shen$^{a,b}$
\\
$^{a}$Shanghai Astronomical Observatory, Chinese Academy of\\
Sciences, Shanghai 200030, People's Republic of China\\
\\
$^{b}$National Astronomical Observatories, Chinese Academy of\\
Sciences, Beijing 100012, People's Republic of China\\
\\
$^{c}$Graduate School of Chinese Academy of Sciences,\\
Beijing 100039, People's Republic of
China\\
\\
\email{fwsu@shao.ac.cn}, \quad \email{ygshen@shao.ac.cn} }

\abstract{ We study analytically quasinormal modes in a wide variety
of black hole spacetimes, including $d$--dimensional asymptotically
flat spacetimes and non-asymptotically flat spacetimes (particular
attention has been paid to the four dimensional case). We extend the
analytical calculation to include first-order corrections to
analytical expressions for quasinormal mode frequencies by making
use of a monodromy technique. All possible type perturbations are
included in this paper. The calculation performed in this paper show
that systematic expansions for uncharged black holes include
different corrections with the ones for charged black holes. This
difference makes them have a different $n$--dependence relation in
the first-order correction formulae. The method applied above in
calculating the first-order corrections of quasinormal mode
frequencies seems to be unavailable for black holes with small
charge. This result supports the Neitzke's prediction. On what
concerns quantum gravity we confirm the view that the $\ln3$ in
$d=4$ Schwarzschild seems to be nothing but some numerical
coincidences. }


\begin{document}

%%%%%%%%%%%%%%%%%%%%%%%%%%%%%%%%%%%%%%%%%%%%%%%%%%%%%%%%%%%%%%%%%
%%%%%%%%%%%%%%%%%%%%%%%%%%%%%%%%%%%%%%%%%%%%%%%%%%%%%%%%%%%%%%%%%

%%%%%%%%%%%%%%%%%%%%%%%%%%%%%%%%%%%%%%%%%%%%%%%%%%%%%%%%%%%%%%%%%
%%%%%%%%%%%%%%%%%%%%%%%%%%%%%%%%%%%%%%%%%%%%%%%%%%%%%%%%%%%%%%%%%

\vfill

\eject

%%%%%%%%%%%%%%%%%%%%%%%%%%%%%%%%%%%%%%%%%%%%%%%%%%%%%%%%%%%%%%%%%
%%%%%%%%%%%%%%%%%%%%%%%%%%%%%%%%%%%%%%%%%%%%%%%%%%%%%%%%%%%%%%%%%

\section{Introduction}

%%%%%%%%%%%%%%%%%%%%%%%%%%%%%%%%%%%%%%%%%%%%%%%%%%%%%%%%%%%%%%%%%
%%%%%%%%%%%%%%%%%%%%%%%%%%%%%%%%%%%%%%%%%%%%%%%%%%%%%%%%%%%%%%%%%

The stability of black holes has been discussed since researchers
found that black holes were shown to radiate and evaporate when we
add in the ideas of quantum mechanics to them\cite{bch, hawking} and
hence people tried to make sure whether the black hole solution
under consideration was really a stable one of the
\textit{classical} theory. The pioneering work on this problem was
carried by Regge and Wheeler\cite{regge-wheeler}, who focused on
analyzing the linear stability of four dimensional Schwarzschild
black hole. They found that one can use a Schr\"odinger--like
equation to describe the linear perturbations. This work was latter
extended to many other black hole solutions and is now known as the
quasinormal modes (QNMs) which can be described as a
``characteristic sound'' of black holes ( A lot of investigation
have been made on this subject\cite{nollert, kokkotas-schmidt,
cardoso phd, sfw}). QNMs are excited by the external
perturbations(may be induced, for example, by the falling matter).
They appear as damped oscillations described by the complex
characteristic frequencies which are entirely fixed by the
parameters of the given black hole spacetime, and independent of the
initial perturbation\cite{Vishveshwara}. These frequencies can be
detected by observing the gravitational wave signal\cite{FE} : this
makes QNMs be of particular relevance in gravitational wave
astronomy.

Although the QNMs are important in the observational aspects of
gravitational waves phenomena mentioned above, there is suggestion
that the asymptotic QNMs may find a very important place in Loop
Quantum Gravity (LQG). Recently Hod \cite{hod} made an interesting
proposal to infer quantum properties of black holes from their
classical oscillation spectrum . The idea was based on the
Bekenstein's conjecture \cite{bekenstein} that in a quantum theory
of gravity the surface area of a non-extremal black hole should have
a discrete eigenvalue spectrum. The eigenvalues of this spectrum are
likely to be uniformly spaced. According to the numerical values
computed by Nollert \cite{nollert} and later confirmed by Andersson
\cite{andersson}, Hod observed that the real parts of the asymptotic
form of high overtones of a Schwarzschild black hole can be written
as :
\begin{equation*}
\frac{\omega_n}{T_H}=(2n+1)\pi i+\ln3,
\end{equation*}
where $T_H$ is the Hawking temperature. Using this conjecture and
Bekenstein's conjecture, he obtained the Bekenstein-Hawking entropy
for the Schwarzschild black hole. He also showed that this approach
is compatible with the statistical mechanical interpretation of
black hole entropy. Later, Dreyer \cite{dreyer} noted that this is
helpful for the  calculation of the so called Barbero-Immirzi
parameter, a free parameter introduced as the Barbero-Immirzi
connection in the calculation of LQG. The only way so far that could
be used to fix this parameter comes from black hole entropy. Dreyer
use the conjecture made by Hod, he fixed the value for the Immirzi
parameter by $\frac{\ln3}{2\pi \sqrt{2}}$. Using this value, he
suggested that ``the appropriate gauge group of quantum gravity is
SO (3) and not its covering group SU (2)''. However, more recently
Corichi\cite{corichi} argued that the LQG allows us to keep SU (2)
as the gauge group, and at the same time have a consistent
description with the results of Dreyer. He reconsidered the physical
process that would give rise to the quasinormal frequency (QN
frequency) as mentioned in\cite{dreyer}: an appearance or
disappearance of a puncture with spin $j_{min}$. Taking into account
of the \textit{local} conservation of fermion number, Corichi
obtained that ``the minimum allowed value for the `spin' of the
resulting free edge is $j_{min}=1$'': this agrees with the results
of Dreyer. However, all this problems are far from being resolved.

As mentioned in the above paragraph, the Hod's conjecture was based
on the numerical results evaluated by Nollert\cite{nollert}. People
may think this agreement is just a coincidence.  Motl\cite{motl,
motl-neitzke}, however, confirmed analytically Nollert's result by
two different methods. In Ref.\cite{motl}, the author used Nollert's
continued fraction expansion for the $4$-dimensional Schwarzschild
and showed that the asymptotic QN frequencies are in good agreement
with Hod's result. The monodromy technique was first introduced in
\cite{motl-neitzke} to analytically compute asymptotic QN
frequencies and later extended in \cite{natario-schiappa}, so that
it can also be used in the computation of $d$--dimensional
asymptotically dS, AdS spacetime. A question, however, produced as
showed in \cite{motl-neitzke,natario-schiappa} is that the
suggestion in \cite{hod, dreyer} was proved not to be universal and
be only applicable to the Schwarzschild solution. It is necessary to
stress that even if the ideas in \cite{hod, dreyer} turn out not to
be universal, it is still the case that QN frequencies will play
some role in the realm of quantum gravity, since the LQG is far from
being successful, and recent studies show that they have
interpretation in conformal field theory through AdS/CFT
correspondence \cite{maldacena,horowitz-hubeny,strominger,fhks}.
Moreover, The possible appearance of $\ln{2}$ in the asymptotic
frequencies\cite{motl-neitzke} could support the claims
\cite{corichi} that the gauge group of LQG should be SU(2) despite
the $\ln{3}$ for Schwarzschild.

It is important to note that the behavior of QNMs for the lowest
modes (frequencies with a smaller imaginary part) is totally
different from the one for the high overtone. We have shown that the
real part of the QN frequencies approaches some certain value as the
imaginary part approaches infinity. There has no such phenomena for
the low-lying modes. A question of particular relevance that
immediately follows is what will happen for the QNMs when the
imaginary part has a middle value between these two extreme cases,
and why these two cases behavior so differently (the resolution of
this problem is helpful to deduce analytically the asymptotic value
of the QN frequencies). Recently Musiri and Siopsis
\cite{musiri-siopsis,musiri-siopsis1} have studied in detail about
this question for Schwarzschild in asymptotically flat and
asymptotically AdS spacetime. Extending the technique introduced in
\cite{motl-neitzke} to obtain a systematic expansion including
corrections in $1/\sqrt{\omega}$, they obtained the $j$ and $l$
dependence of the first correction for arbitrary $j$. Their results
are in good agreement with the results obtained by numerical methods
in the case of scalar and gravitational waves. However, they have
discussed nothing about more general spacetime background, such as
the case of Reissner-Nordstr\"{o}m (RN) black holes and higher
dimensions. The main purpose of this paper is to study analytically
the first-order correction to the asymptotic form of QN frequencies
for more general sapcetime background. In this work we shall make
use of the remarkable results obtained by Ishibashi and Kodama (we
refer the reader to \cite{kodama-ishibashi-1, kodama-ishibashi-2,
kodama-ishibashi-3} for detail on this subject). They studied in
detail the perturbation theory of static, spherically symmetric
black holes in any space-time dimension $d>3$ and allowing for the
possibilities of both electromagnetic charge and a background
cosmological constant. According to them, the perturbations come in
three types: tensor type perturbations, vector type perturbations
and scalar type perturbations, and linear perturbations in
$d$--dimensions can be described by a set of equations which may be
denoted as the Ishibashi--Kodama (IK) master equations.

The organization of this paper is as follows. In next section we
apply the monodromy method introduced in \cite{motl-neitzke} to the
IK master equations expanded near the several singularities in the
complex plane to analytically compute both zeroth-order and
first-order asymptotic quasinormal frequencies for static,
spherically symmetric black hole spacetimes in dimension $d>3$.
Section 3 is the last section of the paper, where we have a
discussion about our results, listing some problems encountered in
this paper. Some future directions are also included in this
section. The last section is our appendices. In appendix A we make
use of expanding the tortoise coordinate to first order at the
singularities in the spacetimes considered in this paper, providing
a full analysis of the potentials at several singularities in the
complex plane, and obtaining a list of first-order IK master
equation potentials.

%%%%%%%%%%%%%%%%%%%%%%%%%%%%%%%%%%%%%%%%%%%%%%%%%%%%%%%%%%%%%%%%%
%%%%%%%%%%%%%%%%%%%%%%%%%%%%%%%%%%%%%%%%%%%%%%%%%%%%%%%%%%%%%%%%%

\section{Perturbative Calculation of Quasinormal Modes}

%%%%%%%%%%%%%%%%%%%%%%%%%%%%%%%%%%%%%%%%%%%%%%%%%%%%%%%%%%%%%%%%%
%%%%%%%%%%%%%%%%%%%%%%%%%%%%%%%%%%%%%%%%%%%%%%%%%%%%%%%%%%%%%%%%%
In this section we first review the perturbation theory roughly for
spherically symmetric, static $d$--dimensional black holes ($d>3$),
with mass $M$, charge $Q$ and background cosmological constant
$\Lambda$, and the computation of QNMs and QN frequencies.
We refer the reader to \cite{natario-schiappa} for more detail. \\
\hspace*{7.5mm}For a massless, uncharged, scalar field, $\Psi$,
after a harmonic decomposition of the scalar field as $\Psi =
\sum_{\ell,m} r^{\frac{2-d}{2}}\ \psi_{\ell} (r,t)\ Y_{\ell m}
\left(  \theta_{i} \right)$, where the $ \theta_{i}$ are the $(d-2)$
angles and the $Y_{\ell m} (  \theta_{i} )$ are the $d$--dimensional
spherical harmonics, and a Fourier decomposition of the scalar field
$\psi_{\ell} (r,t) = \Psi (r) e^{i \omega t}$ , the wave equation
can be decoupled as a Schr\"odinger--like equation

\begin{equation} \label{schrodinger}
- \frac{ d^{2} \Psi (r_*)}{dr_*^{2}} + V (r_*) \Psi (r_*) =
\omega^{2} \Psi (r_*),
\end{equation}

\noindent where $r_*$ is tortoise coordinate defined as $dr_* =
\frac{dr}{f(r)}$ and $V(r_*)$ is the potential, both determined from
the function $f(r)$ in the background metric. The potential $V(r_*)$
depends on the background space-time metric and the perturbative
type (appendix \ref{appendixA}). For QNMs we need some boundary
conditions, so that
\begin{eqnarray*}
\Psi (r_*) &\sim& e^{i\omega r_*}\;\, {\mathrm{as}}\;\, r_* \to - \infty, \\
\Psi (r_*) &\sim& e^{-i\omega r_*}\;\, {\mathrm{as}}\;\, r_* \to +
\infty.
\end{eqnarray*}

Using this boundary conditions and the monodromy technique, we shall
show how to calculate the asymptotic QN frequencies and their
first-order correction in all static, spherically symmetric black
hole spacetimes (including asymptotically flat spacetimes and
non-asymptotically flat spacetimes). As an example, we may pay more
attention to the case of $d=4$. For some cases, we shall list some
corrected QN frequencies, so that we can show it is a reasonable
correction by comparing with the numerical results.

%%%%%%%%%%%%%%%%%%%%%%%%%%%%%%%%%%%%%%%%%%%%%%%%%%%%%%%%%%%%%%%%%

\subsection{The Schwarzschild Case}

%%%%%%%%%%%%%%%%%%%%%%%%%%%%%%%%%%%%%%%%%%%%%%%%%%%%%%%%%%%%%%%%%

Although the perturbative calculation for the Schwarzschild solution
in $4$--dimension and higher dimension have been discussed in
\cite{musiri-siopsis} and \cite{cly}, respectively, we first review
it roughly for completeness.

 For Schwarzschild black hole, we have
$$
f(r)=1-\frac{2 m}{r^{d-3}},
$$
with the roots
$$
r_n=\left| (2m)^{\frac1{d-3}}\right| \exp \left(\frac{2\pi i}{d-3}
n\right), \ \ \ \ n =0, 1, \cdots, d-4.
$$
The radial wave equation for gravitational perturbations in the
black-hole background can be written as
$$
- \frac{ d^{2} \Psi(r_*)}{dr_*^{2}}  + V \big[ r(r_*) \big] \Psi
(r_*) = \omega^{2} \Psi (r_*)
$$

\noindent in the complex $r$--plane.

As mentioned above, the boundary conditions are
$$
\Psi (r_*) \sim e^{\mp i\omega r_*}\;\, {\mathrm{as}}\;\, r_* \to
\pm \infty,
$$
assuming ${\mathbb{R}}{\mathrm{e}}\, \omega > 0$. Then we obtain
\begin{eqnarray}\label{bounder}
\nonumber F (r_*) &\sim& 1\;\, {\mathrm{as}}\;\, r_* \to + \infty, \\
F (r_*) &\sim& e^{2i\omega r_*}\;\, {\mathrm{as}}\;\, r_* \to -
\infty,
\end{eqnarray}
if we rewrite the QNMs as $F(r_*)=e^{i \omega r_*}\Psi(r_*)$. The
clockwise monodromy of $F(r_*)$ around the $r=r_H$ can be easily
obtained by continuing the coordinate $r$ analytically into the
complex plane, \textit{i.e.},
$$
\mathcal{M}(r_H)=e^{\frac{2\pi \omega}{k_H}},
$$
where $k_H= \frac12 f'(r_H) $ is the surface gravity at the horizon.

Near the black hole singularity ($r \sim 0$), the tortoise
coordinate may be expanded as
$$
r_*=\int \frac{dr}{f(r)}=-\frac{1}{d-2}\frac{r^{d-2}}{2
m}-\frac{1}{2d-5}\frac{r^{2d-5}}{(2 m)^2}+\cdots,
$$
\noindent where $f(r)=1-\frac{2 m}{r^{d-3}}$ and $ m$ is the mass of
the black hole\footnote{In fact, one can relate it with the ADM mass
$M$ by
$$
M = \frac{\left( d-2 \right) {\mathcal{A}}_{d-2}}{8\pi G_{d}}  m,
$$
where ${\mathcal{A}}_{n}$ is the area of an unit $n$--sphere, $
{\mathcal{A}}_{n} = \frac{2\pi^{\frac{n+1}{2}}}{\Gamma \left(
\frac{n+1}{2} \right)}$.}. When we define $z=\omega r_*$, the
potential near the black hole singularity for the three different
type perturbations can be expanded , respectively, as (appendix
\ref{appendixA})
\begin{equation}\label{potential-sw}
V [ z ] \sim -\frac{\omega^2}{4z^2} \left \{1-j^2-W(j)
\left(\frac{z}{\omega}\right)^{(d-3)/(d-2)}+\cdots \right\} ,
\end{equation}
\noindent where
\begin{eqnarray*}
W(j)=
 \begin{cases}
  W_{ST} &\text{$j=0$},\\
  W_{SV} &\text{$j=2$},\\
  W_{SS} &\text{$j=0$},
 \end{cases}
\end{eqnarray*}
and the explicit expressions of $W_{ST}$, $W_{SV}$ and $W_{SS}$ can
be found in appendix \ref{appendixA}. Then the Schr\"{o}dinger-like
wave equation (\ref{schrodinger}) with the potential
(\ref{potential-sw}) wave equation can be depicted as
\begin{equation}\label{waveequations}
 \left(\mathcal{H}_0
+\omega^{-\frac{d-3}{d-2}}\mathcal{H}_1\right) \Psi=0,
\end{equation}
  where
$\mathcal{H}_0$ and $\mathcal{H}_1$ are defined as
\begin{eqnarray*}
\mathcal{H}_0=\frac{d^2}{dz^2}+\left[\frac{1-j^2}{4z^2}+1\right] , \
\ \ \mathcal{H}_1=-\frac{W(j)}{4}z^{-\frac{d-1}{d-2}}.
\end{eqnarray*}
Taking into account of $\omega\rightarrow\infty$, the zeroth-order
wave equation becomes
$$
\mathcal{H}_0 \Psi^{(0)}=0,
$$
with general solutions in the form of
\begin{equation}\label{solution}
\Psi^{(0)}=A_+J_{j/2}(z)+A_-J_{-j/2}(z),
\end{equation}
where and below $J_\nu(x)$ represents a Bessel function of the first
kind. According to the boundary conditions (\ref{bounder}), one can
define
$$
F^{(0)}(z)=F_+^{(0)}(z)-e^{-\pi j i/2}F_-^{(0)}(z),
$$
which approaches $-e^{-i\alpha_+}\sin \frac{j\pi}{2}$ as $z
\rightarrow +\infty$, where $\alpha_{\pm}=\frac{(1\pm j)\pi}{4}$.
Considering the behavior of the wave function as
$z\rightarrow-\infty$, we may deduce that
$$
\mathcal{M}(r_H) \sim -\frac{\sin3j\pi/2}{\sin j\pi/2},
$$
leading to the generic expression of $d$-dimensional QN frequencies
as showed in \cite{natario-schiappa}
$$
\frac{\omega}{T_H}=(2n+1)\pi i +\ln(1+2\cos j\pi),
$$
where $T_H$ is the Hawking temperature.

Next we calculate the first-order correction of the Schwarzschild
black hole spacetimes. We first expand the wave function to the
first order in $1/\omega^{(d-3)/(d-2)}$ as
$$
\Psi=\Psi^{(0)}+\frac 1 {\omega^{(d-3)/(d-2)}} \Psi^{(1)}.
$$
Then one can rewrite Eq. (\ref{waveequations}) as
\begin{equation}\label{waveequation}
 \mathcal{H}_0
\Psi^{(1)}+\mathcal{H}_1 \Psi^{(0)}=0.
\end{equation}
The general solution of Eq. (\ref{waveequation}) is
\begin{equation}\label{solution1}
\Psi_{\pm}^{(1)}=\mathcal{C}\Psi_+^{(0)}\int_0^z
\Psi_-^{(0)}\mathcal{H}_1
\Psi_{\pm}^{(0)}-\mathcal{C}\Psi_-^{(0)}\int_0^z
\Psi_+^{(0)}\mathcal{H}_1 \Psi_{\pm}^{(0)},
\end{equation}
where $\mathcal{C}=-\frac{1}{\sin j\pi /2}$, and the wave function
$\Psi_{\pm}^{(0)}$ are
\begin{equation*}
\Psi_{\pm}^{(0)}=\sqrt{\frac{\pi z}{2}}J_{\pm j/2}(z).
\end{equation*}
Taking into consideration both the boundary conditions
(\ref{bounder}) and the behavior of wave function $F(z)$ as
$z\rightarrow \pm\infty$ along the real axis, one may deduce
$$
\mathcal{M}(r_H) \sim -\frac{\sin3j\pi/2}{\sin
j\pi/2}\left(1+\frac{\xi_--K}{\omega^{(d-3)/(d-2)}}\right),
$$
leading to the generic expression of $d$-dimensional QN frequencies
expressed as
$$
\frac{\omega}{T_H}=(2n+1)\pi i +\ln(1+2\cos
j\pi)+\frac{corr_d}{(n+1/2)^{(d-3)/(d-2)}},
$$
where we have defined $\xi_{\pm}=c_{\pm\pm}e^{\pm ij\pi/2}-c_{+-}$,
and
$$
c_{\pm\pm}=\mathcal{C}\int_0^{\infty}\Psi_{\pm}^{(0)}\mathcal{H}_1
\Psi_{\pm}^{(0)}.
$$
And $K$ is defined as
$$
K=e^{i\pi(3+6\eta)/4}\left[\xi_{+}\sin(\frac{6\eta-3}{4}\pi)+i\xi_-\cos(\frac{6\eta-3}
{4}\pi)-\xi\cos(\frac{6\eta-3}{4}\pi) \cot \frac{3j\pi}{2}\right],
$$
where $\eta=(d-4)/2(d-2)$ and $\xi=\xi_+ +\xi_-$\footnote{Throughout
this work we have the same definitions for $\alpha_{\pm}$,
$c_{\pm}$, $\xi_{\pm}$, $\xi$, and $\mathcal{C}$.}.

For $d=4$, we obtain the same result as \cite{musiri-siopsis}
$$
corr_4=(1-i)\frac{3l(l+1)+1-j^2}{24\sqrt{2}(\pi)^{3/2}}\frac{\sin(2j\pi)}{\sin(3j\pi/2)}
\Gamma^2 (1/4)\Gamma(1/4+j/2)\Gamma(1/4-j/2).
$$
The results above are shown to be dimension dependent and related
closely to $l$ and $j$. It is reasonable that these $d$--dimensional
frequencies would indicate some information about back ground
spacetime and perturbation types as shown in lowest QNMs. It
approaches a constant for the real part of the QN frequency once we
let $n\rightarrow \infty$ as expected in the
literature\cite{motl,motl-neitzke, konoplya-1}. In addition,
\cite{musiri-siopsis} showed that the results do agree with the
result from a WKB analysis\cite{amdb}, as well as the numerical
results \cite{nollert} for scalar perturbations and gravitational
perturbations, respectively.

%%%%%%%%%%%%%%%%%%%%%%%%%%%%%%%%%%%%%%%%%%%%%%%%%%%%%%%%%%%%%%%%%

\subsection{The Reissner--Nordstr\"om Case}

%%%%%%%%%%%%%%%%%%%%%%%%%%%%%%%%%%%%%%%%%%%%%%%%%%%%%%%%%%%%%%%%%
Now we discuss the QNMs of the Reissner--Nordstr\"om
$d$--dimensional black hole including first-order corrections. The
calculation for zeroth-order was first done in \cite{motl-neitzke},
and latter in \cite{natario-schiappa}. In order to state more
clearly, we start with the zeroth-order calculation. For
Reissner--Nordstr\"om black hole, we have
$$
f(r)=1-\frac{2 m}{r^{d-3}}+\frac{ q^2}{r^{2d-6}},
$$
with the roots
$$
r_n^{\pm}=\left| \left( m\pm\sqrt{ m^2-
q^2}\right)^{\frac1{d-3}}\right| \exp \left(\frac{2\pi i}{d-3}
n\right), \ \ \ \ n =0, 1, \cdots, d-4,
$$
where $ q$ is the charge of the black hole\footnote{One can relate
it with the charge $Q$ by
$$
Q^2 = \frac{(d-2)(d-3) q^2}{8\pi G_d}.
$$
}.

Again we define
$$
F(r_*)=e^{i \omega r_*}\Psi(r_*).
$$
Then the clockwise monodromy of $F(r_*)$ around the outer horizon
$r=r_+$ can be obtained by continuing the coordinate $r$
analytically into the complex plane, \textit{i.e.},
$$
\mathcal{M}(r_+)=e^{\frac{2\pi \omega}{k_+}},
$$
where $k_+= \frac12 f'(r_+) $ is the surface gravity at the outer
horizon.

Near the black hole singularity ($r \sim 0$), the tortoise
coordinate may be expanded as
\begin{equation}\label{r0rn}
r_*=\int \frac{dr}{f(r)}=\frac{1}{2d-5}\frac{r^{2d-5}}{ q^2}+\frac{2
m}{3d-8}\frac{r^{3d-8}}{ q^4}+\cdots.
\end{equation}
One can easily learn from (\ref{r0rn}) that $r_* \rightarrow \infty$
as $ q \rightarrow 0$. In fact, in our procedure for expanding
(\ref{r0rn}), we have assumed $\frac{r}{r_0^-}\ll 1$. As a result,
instead of expanding potential to the first order in
$1/\omega^{(d-3)/(2d-5)}$, we must expand it in
$1/\left[(r_0^-)^{2d-5}\omega\right]^{(d-3)/(2d-5)}$. After defining
$z=\omega r_*$, the potential for the three different type
perturbations can be then expanded, respectively, as (appendix
\ref{appendixA})
\begin{equation}\label{potential-rn}
V [ z ] \sim -\frac{\omega^2}{4z^2} \left \{1-j^2-W(j)
\left(\frac{z}{(r_0^-)^{2d-5}\omega}\right)^{(d-3)/(2d-5)}+\cdots
\right\} ,
\end{equation}
where
\begin{eqnarray*}
W(j)=
 \begin{cases}
  W_{RNT} &\text{$j=j_T$},\\
  W_{RNV^{\pm}} &\text{$j=j_{V^{\pm}}$},\\
  W_{RNS^{\pm}} &\text{$j=j_{S^{\pm}}$},
 \end{cases}
\end{eqnarray*}
 and the explicit expressions of $W_{RNT}$, $W_{RNV^{\pm}}$ and $W_{RNS^{\pm}}$ can be found in appendix \ref{appendixA}.
Then the Schr\"{o}dinger-like wave equation (\ref{schrodinger}) with
the potential (\ref{potential-rn}) can be depicted as
\begin{equation}\label{waveequation2}
 \left(\mathcal{H}_0
+\left[(r_0^-)^{2d-5}\omega\right]^{-\frac{d-3}{2d-5}}\mathcal{H}_1\right)
\Psi=0,
\end{equation}
 where
$\mathcal{H}_0$ and $\mathcal{H}_1$ are defined as
\begin{eqnarray*}
\mathcal{H}_0=\frac{d^2}{dz^2}+\left[\frac{1-j^2}{4z^2}+1\right] , \
\ \ \mathcal{H}_1=-\frac{W(j)}{4}z^{-\frac{3d-7}{2d-5}}.
\end{eqnarray*}

\FIGURE[ht]{\label{StokesRN}
    \centering
    \psfrag{A}{$A$}
    \psfrag{B}{$B$}
    \psfrag{r0}{$r_0^-$}
    \psfrag{r1}{$r_1^-$}
    \psfrag{r2}{$r_2^-$}
    \psfrag{R0}{$r_0^+$}
    \psfrag{R1}{$r_1^+$}
    \psfrag{R2}{$r_2^+$}
    \psfrag{Re}{$\re$}
    \psfrag{Im}{$\im$}
    \psfrag{contour}{contour}
    \psfrag{Stokes line}{Stokes line}
    \epsfxsize=.6\textwidth
    \leavevmode
    \epsfbox{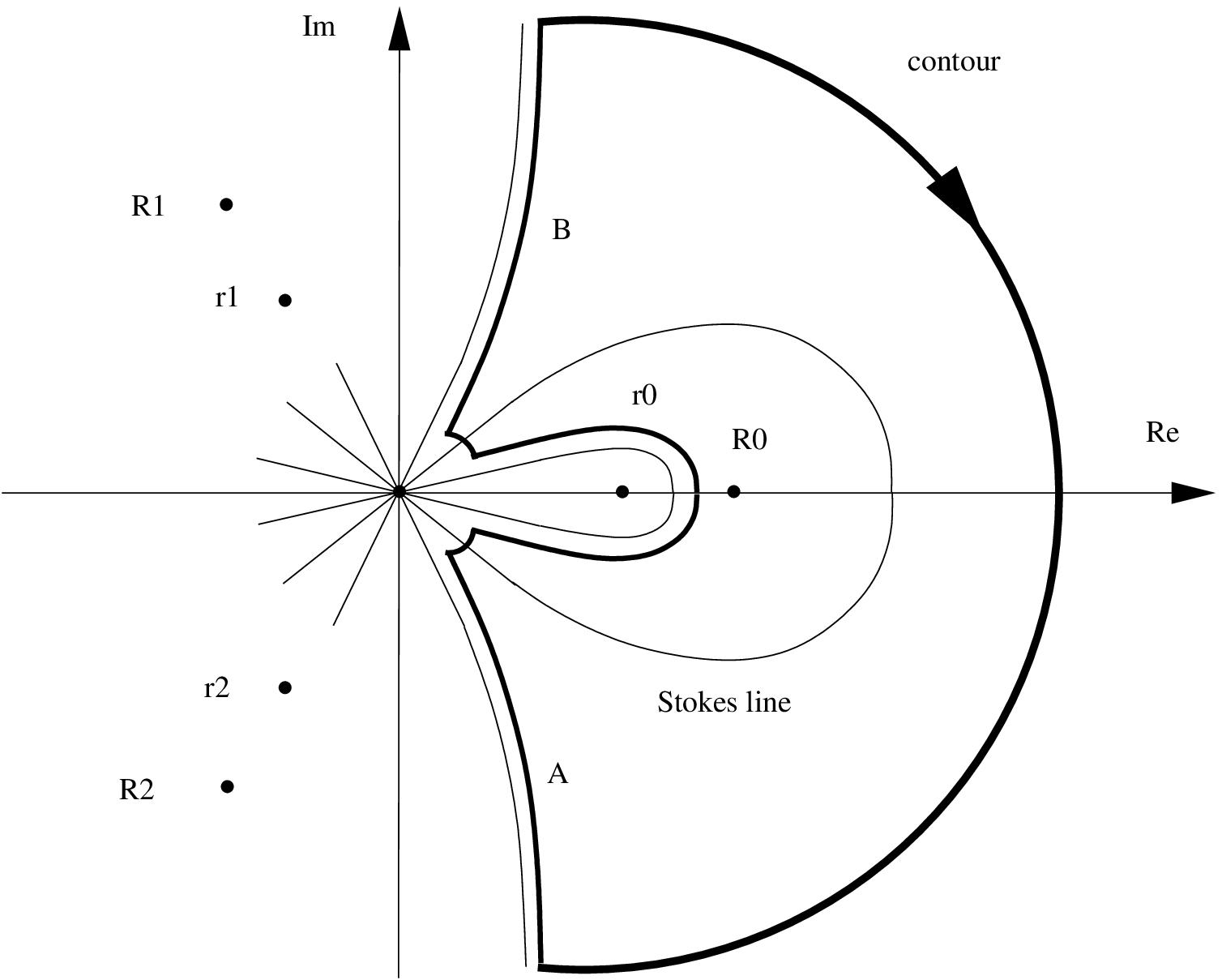}
\caption{Stokes line for the Reissner--Nordstr\"om black hole, along
with the chosen contour for monodromy matching, in the case $d=6$ (
we refer the reader to \cite{natario-schiappa} for detail, and a
more complete list of figures in dimensions $d=4$, $d=5$, $d=6$ and
$d=7$ ).} } Taking into account of $\omega\rightarrow\infty$, the
zeroth-order wave equation becomes
$$
\mathcal{H}_0 \Psi^{(0)}=0,
$$
with general solutions in the form of Eq. (\ref{solution}).
According to the boundary conditions (\ref{bounder}), one can define
$$
F^{(0)}(z)=F_+^{(0)}(z)-e^{-\pi j i/2}F_-^{(0)}(z),
$$
which approaches $-e^{-i\alpha_+}\sin \frac{j\pi}{2}$ as $z
\rightarrow +\infty$. This holds at point $A$ in Figure
\footnote{All figures in this paper are provided by J. Nat\'{a}rio
and R. Schiappa in \cite{natario-schiappa}.} \ref{StokesRN}. By
going around an arc of angle of $\frac{2\pi}{2d-5}$ in complex
$r$--plane (rotating from $A$ to the next branch), $z$ rotates
through an angle of $2\pi$ in $z$--plane, leading to the wave
function
\begin{equation}\label{rnf1}
F^{(0)}(e^{2\pi i}z)=\frac1 {2}
\left[e^{2iz}\left(e^{3i\alpha_+}-e^{-ij\pi/2}e^{3i\alpha_-}\right)+
\left(e^{5i\alpha_+}-e^{-ij\pi/2}e^{5i\alpha_-}\right)\right].
\end{equation}
As one follows the contour around the inner horizon $r=r_-$, the
wave function will be of the form
\begin{equation*}
F^{(0)}(z-\delta)=A_+F_+^{(0)}(z-\delta)-A_-F_-^{(0)}(z-\delta),
\end{equation*}
where
$$
\delta = \frac{2\omega\pi i}{f'(r_-)} = \frac{\omega\pi i}{k_-},
$$
and $ k_- = \frac12 f'(r_-) $ is the surface gravity at the inner
horizon. Notice that $z-\delta< 0$ on this branch, one can easily
obtain
\begin{equation}\label{rnf3}
F^{(0)}(z-\delta)=\frac1 {2}
\left[e^{2i(z-\delta)}\left(A_+e^{i\alpha_+}-A_-e^{i\alpha_-}\right)+
\left(A_+e^{-i\alpha_+}-A_-e^{-i\alpha_-}\right)\right],
\end{equation}
where we have used
\begin{equation*}
J_\nu(z)=\sqrt{\frac{2}{\pi z}} \cos\left(z+\frac{\nu
\pi}{2}+\frac{\pi}4 \right) , \qquad z \ll -1.
\end{equation*}
Since on the same branch, we must let
$F^{(0)}(z-\delta)=F^{(0)}(e^{2\pi i}z)$, and hence it is easily
seen from Eqs.(\ref{rnf1}) and (\ref{rnf3}) that we must have
\begin{eqnarray*}
&& e^{3i\alpha_+} - e^{-ij\pi/2}e^{3i\alpha_-} = A_+  e^{i\alpha_+} e^{-2i\delta} - A_- e^{i\alpha_-} e^{-2i\delta},\\
&& e^{5i\alpha_+} - e^{-ij\pi/2}e^{5i\alpha_-} = A_+ e^{-i\alpha_+}
- A_- e^{-i\alpha_-}.
\end{eqnarray*}
Then we obtain
\begin{eqnarray}\label{AA1}
\nonumber && A_+=2e^{2i\delta}e^{\frac {\pi i}2} \cos\frac{j \pi}2-e^{i\pi\frac{3-j}2}\frac{\sin (3j\pi/2)}{\sin (j\pi/2)},\\
&& A_-=2e^{2i\delta}e^{i\pi\frac {1-j}2} \cos\frac{j
\pi}2-e^{\frac{3i\pi}2}\frac{\sin (3j\pi/2)}{\sin (j\pi/2)}.
\end{eqnarray}
Finally, we must rotate to the branch containing point $B$. This
makes $z - \delta$ rotate through an angle of $2\pi$, leading to the
wave function
\begin{equation*}
F^{(0)}(e^{2\pi i}(z-\delta))=2 e^{2i\delta}e^{-i\alpha_+}\cos
(\frac{j\pi}2)\sin(j\pi) +e^{-i\alpha_+}\sin (\frac{3j\pi}2).
\end{equation*}
Consequently, we deduce that
\begin{equation}\label{rnzero}
e^{\frac{2\pi\omega}{k_+}}=-2e^{2i\delta}
[1+\cos(j\pi)]-(1+2\cos(j\pi)),
\end{equation}
which agrees with the result shown in \cite{natario-schiappa}.

Next we compute the first-order correction of the asymptotic QNMs of
the $d$--dimensional RN black hole using the monodromy method. The
same reason as mentioned above, we must expand the wave function to
the first order in
$1/\left[(r_0^-)^{2d-5}\omega\right]^{(d-3)/(2d-5)}$ as
$$
\Psi=\Psi^{(0)}+\frac 1
{\left[(r_0^-)^{2d-5}\omega\right]^{(d-3)/(2d-5)}} \Psi^{(1)}.
$$
Then one can rewrite Eq. (\ref{waveequation2}) as
\begin{equation*}
 \mathcal{H}_0
\Psi^{(1)}+\mathcal{H}_1 \Psi^{(0)}=0.
\end{equation*}
The general solution of this equation, as mentioned in the last
subsection, is
\begin{equation}\label{solution2}
\Psi_{\pm}^{(1)}=\mathcal{C}\Psi_+^{(0)}\int_0^z
\Psi_-^{(0)}\mathcal{H}_1
\Psi_{\pm}^{(0)}-\mathcal{C}\Psi_-^{(0)}\int_0^z
\Psi_+^{(0)}\mathcal{H}_1 \Psi_{\pm}^{(0)}.
\end{equation}
The behavior as $z \gg 1$ is found to be
\begin{equation*}
\Psi_{\pm}^{(1)}(z) \sim c_{- \pm}\cos (z-\alpha_+)-c_{+ \pm}\cos
(z-\alpha_-).
\end{equation*}
After defining
\begin{eqnarray*}
\Psi &=& \Psi_+^{(0)} +\frac
1{\left[(r_0^-)^{2d-5}\omega\right]^{(d-3)/(2d-5)}}\Psi_+^{(1)}-\\
&&-e^{-ij\pi/2}\left(1-\frac{\xi}{\left[(r_0^-)^{2d-5}\omega\right]^{(d-3)/(2d-5)}}\right)\left(\Psi_-^{(0)}
+\frac
1{\left[(r_0^-)^{2d-5}\omega\right]^{(d-3)/(2d-5)}}\Psi_-^{(1)}
\right),
\end{eqnarray*}
and using boundary condition (\ref{bounder}), one finds that the
wave function $F(z)$ approaches
$$
 F(z) \sim
-e^{-i \alpha_+} \sin \frac{j \pi}{2}
\left[1-\frac{\xi_-}{\left[(r_0^-)^{2d-5}\omega\right]^{(d-3)/(2d-5)}}\right],
$$
as $z \rightarrow \infty$.\\
 The function $\Psi_{\pm}^{(1)}$ defined
in (\ref{solution2}) follows that
$$
\Psi_{\pm}^{(1)}=z^{1\pm j/2+\eta}G_{\pm}(z),
$$
where $\eta=-1/(4d-10)$, and $G_{\pm}(z)$ are even analytic
functions of $z$. By going around an arc of angle of
$\frac{2\pi}{2d-5}$ in complex $r$--plane (rotating from $A$ to the
next branch), $z$ rotates through an angle of $2\pi$ in $z$--plane,
leading to the wave function
\begin{equation*}
F(e^{2\pi i}z)=F^{(0)}(e^{2\pi i}z) + \frac{
e^{i\pi(2+j+2\eta)}}{\left[(r_0^-)^{2d-5}\omega\right]^{(d-3)/(2d-5)}}\left[F_+^{(1)}(z)-e^{-5j\pi
i/2}\left(F_-^{(1)}(z)-\xi e^{-i\pi(1+2\eta)}
F_-^{(0)}(z)\right)\right].
\end{equation*}
As one follows the contour around the inner horizon $r=r_-$, the
wave function will be of the form
\begin{equation*}
F(z-\delta)=\tilde{A}_+F_+(z-\delta)-\tilde{A}_-F_-(z-\delta),
\end{equation*}
where again
$$
\delta = \frac{2\omega\pi i}{f'(r_-)} = \frac{\omega\pi i}{k^-},
$$
and $ k^- = \frac12 f'(r_-) $ is the surface gravity at the inner
horizon. For highly damped QNMs ($\omega \gg 1$), approximately, we
have
$$
\tilde{A}_+ =A_+, \ \ \ \ \tilde{A}_- =A_-.
$$
Finally, we must rotate to the branch containing point $B$. This
makes $z - \delta$ rotate through an angle of $2\pi$, leading to the
wave function
\begin{eqnarray*}
F(e^{2\pi
i}(z-\delta))&=& e^{2i(z-\delta)}B+\frac12\left(A_+e^{3i\alpha_+}-A_-e^{3i\alpha_-}\right)\\
&&+ \frac{e^{2\eta \pi
i}}{2\left[(r_0^-)^{2d-5}\omega\right]^{(d-3)/(2d-5)}}
\left[A_+\xi_+e^{3i\alpha_+}+A_-\xi_-e^{3i\alpha_-}+A_-\xi e^{-2\eta
\pi i}e^{3i\alpha_-}\right],
\end{eqnarray*}
where $B$ is a constant which is not needed in our calculation
because $e^{2i(z-\delta)}\rightarrow 0$ as $z\rightarrow-\infty$.
Consequently, as one uses the explicit expression of $A_+$ and $A_-$
in Eq. (\ref{AA1}), we deduce that
\begin{eqnarray*}
F(e^{2\pi
i}(z-\delta))&=&2e^{2i\delta}e^{-i\alpha}\cos(\frac{j\pi}2)\sin(j\pi)\left(
1-\frac{K_1}{\left[(r_0^-)^{2d-5}\omega\right]^{(d-3)/(2d-5)}}\right)\\
&&+e^{-i\alpha}\sin(\frac {3j\pi}2)\left(
1-\frac{K_2}{\left[(r_0^-)^{2d-5}\omega\right]^{(d-3)/(2d-5)}}\right),
\end{eqnarray*}
where
\begin{eqnarray*}
K_1&=& e^{i\pi (\eta+3/2)}[\xi_+\sin(\eta
\pi)-\xi\cos(\eta\pi)\cot(j\pi)+i\xi_-\cos(\eta\pi)],\\
K_2&=& e^{i\pi (\eta+3/2)}[\xi_+\sin(\eta
\pi)-\xi\cos(\eta\pi)\cot(j\pi/2)+i\xi_-\cos(\eta\pi)].
\end{eqnarray*}
Finally, we have
\begin{equation*}
e^{\frac{2\pi\omega}{k_+}}=-2e^{2i\delta}
[1+\cos(j\pi)]\left(1+\frac{\xi_--K_1}{\left[(r_0^-)^{2d-5}\omega\right]^{(d-3)/(2d-5)}}\right)
-[1+2\cos(j\pi)]\left(1+\frac{\xi_--K_2}{\left[(r_0^-)^{2d-5}\omega\right]^{(d-3)/(2d-5)}}\right),
\end{equation*}
leading to the generic expression of $d$-dimensional QN frequencies
$$
e^{\frac{\omega}{T_H^+}}+2e^{-\frac{\omega}{T_H^-}}
[1+\cos(j\pi)]\left(1+\frac{\xi_--K_1}{\left[(r_0^-)^{2d-5}\omega\right]^{(d-3)/(2d-5)}}\right)
+[1+2\cos(j\pi)]\left(1+\frac{\xi_--K_2}{\left[(r_0^-)^{2d-5}\omega\right]^{(d-3)/(2d-5)}}\right)=0,
$$
where $T_H^+$ and $T_H^-$ represent the Hawking temperature at the
outer and inner
horizons (notice that $T_H^-<0$).\\
For $d=4$, $j_T=j_{S^{\pm}}=1/3$, $j_{V^{\pm}}=5/3$, one obtains
\begin{equation*}
e^{\frac{\omega}{T_H^+}}+3e^{-\frac{\omega}{T_H^-}}
\left(1-\frac{(\sqrt{3}-i)\mathcal{E}_1}{\left[(r_0^-)^{3}\omega\right]^{1/3}}\right)
+2\left(1-\frac{(\sqrt{3}-i)\mathcal{E}_2}{\left[(r_0^-)^{3}\omega\right]^{1/3}}\right)=0,
\end{equation*}
where
$$
\mathcal{E}_1=\frac{1+\sqrt{3}i}{4}[\xi_+-\xi_-+\sqrt{3}\cot(j\pi)],\
\ \
\mathcal{E}_2=\frac{1+\sqrt{3}i}{4}[\xi_+-\xi_-+\sqrt{3}\cot(j\pi/2)].
$$
In the case of $ q\ll  m$, one has $-\frac1{T_H^-}\rightarrow 0$,
which leads to
$$
\frac{\omega}{T_H^+}\sim (2n+1)\pi
i+\log5-\frac{(\sqrt{3}-i)(\mathcal{E}_1+\mathcal{E}_2)}{5r_0^-[(8n+4)k_+]^{1/3}}.
$$
In the case of $ q \rightarrow  m$, one has $\frac1{T_H^+}\approx
-\frac1{T_H^-}$, which leads to
$$
\frac{\omega}{T_H^+}\sim (2n+1)\pi
i-\log2+\frac{(\sqrt{3}-i)(\mathcal{E}_1-4\mathcal{E}_2)}{4r_0^-[(8n+4)k_+]^{1/3}}.
$$
In other cases, approximately, one has
\begin{equation}\label{corrrn}
e^{\frac{\omega}{T_H^+}}+3e^{-\frac{\omega}{T_H^-}}
\left(1-\frac{(\sqrt{3}-i)\mathcal{E}_1}{r_0^-[(8n+4)k_+]^{1/3}}\right)
+2\left(1-\frac{(\sqrt{3}-i)\mathcal{E}_2}{r_0^-[(8n+4)k_+]^{1/3}}\right)=0,
\end{equation}
 In order to obtain an
explicit expression of $\mathcal{E}_1$ and $\mathcal{E}_2$, we need
the integral
$$\mathcal{J}(\nu, \mu)\equiv \int_0^{\infty}dz z^{-2/3}
J_{\nu}(z)J_{\mu}(z)=\frac{\Gamma(\frac23)\Gamma(\frac{
\mu+\nu+1/3}2)}{\sqrt[3] {4}\Gamma(\frac{\nu-\mu+5/3}2)\Gamma(\frac{
\mu+\nu+5/3}2)\Gamma(\frac{ \mu-\nu+5/3}2)}.
$$
As a result, we obtain
\begin{eqnarray*}
\xi_+ &=& \frac{\pi W(j)}{8\sin(j\pi/2)}\left[\mathcal{J}(j/2,j/2)
e^{ij\pi/2} -
\mathcal{J}(j/2,-j/2)\right],\\
\xi_- &=& \frac{\pi W(j)}{8\sin(j\pi/2)}\left[\mathcal{J}(-j/2,-j/2)
e^{-ij\pi/2} - \mathcal{J}(j/2,-j/2)\right],
\end{eqnarray*}
Immediately, one has
\begin{eqnarray*}
\mathcal{E}_1 &=&
-\frac{(\sqrt{3}-i)W{(j)}}{64\sqrt[3]{4}\pi^2}\times\frac{\cos
(3j\pi/2)}{\cos^2(j\pi/2)}\cdot\Gamma(2/3)\Gamma(1/6+j/2)
\Gamma(1/6-j/2)\Gamma^2(1/6),\\
\mathcal{E}_2 &=&
-\frac{(\sqrt{3}-i)W{(j)}}{32\sqrt[3]{4}\pi^2}\cdot
\Gamma(2/3)\Gamma(1/6+j/2) \Gamma(1/6-j/2)\Gamma^2(1/6).
\end{eqnarray*}

In this way, one can easily obtain the value of QN frequencies in
Eq. (\ref{corrrn}). For tensor type perturbations, $W_{RNT}=0$, so
$\mathcal{E}_1=\mathcal{E}_2=0$, leading to the same results as the
zeroth-order asymptotic QN frequencies. However, it seems
unavailable for scalar type perturbations, since in this case we
have $j\rightarrow1/3$, which may induce the integral
$\mathcal{J}(-1/3,-1/3)$ approaches infinity. It is interesting to
investigate this problem in detail. Is there any other methods can
avoid this singularity? For vector type perturbations , we have
$j\rightarrow5/3$, which lead to
\begin{eqnarray*}
\mathcal{E}_1 &\sim& 0,\\
\mathcal{E}_2 &\sim& (0.3734 -
0.2156i)\left(1\pm\sqrt{9+4\ell(\ell+1) q^2-8 q^2}\right)(1-\sqrt{1-
q^2}) q^{-4/3},
\end{eqnarray*}
with $ m=1$. From here we find: (1) in order to insure
$r_0^-[(8n+4)k_+]^{1/3}\gg 1$ as one calculates the QN frequencies
of the first-order correction, the imaginary part of the frequencies
( or the modes $n$) needs bigger values for a black hole with small
charge, since $r_0^- \sim  q^2\rightarrow 0$ as $ q\rightarrow0$.
This confirms the prediction made by Neitzke in \cite{neitzke}: the
required $n$ diverges as $ q\rightarrow 0$, and the corrections
would blow up this divergence; (2) just like the case of
zeroth-order QN frequencies, the $ q\rightarrow0$ limit of RN
$corr_4$ does not yield Schwarzschild $corr_4$, and the same thing
happens in the limit of $ q^2\rightarrow  m^2$, as we shall see in
the following; (3) the first-order correction to the asymptotic QN
frequencies are shown to be dimension dependent and related closely
to $\ell$, $j$, and the charge $ q$.

QN frequencies of RN black holes were calculated numerically by
Berti and Kokkotas in \cite{berti-kokkotas-2}. They found that very
highly damped QNMs of RN black holes have an oscillatory behavior as
a function of the charge. Their results have a good agreement with
the zeroth-order analytical formula (\ref{rnzero}). However, it is
necessary to perform further checks to our first-order corrected
results both in four and higher dimensional black hole spacetime.

%%%%%%%%%%%%%%%%%%%%%%%%%%%%%%%%%%%%%%%%%%%%%%%%%%%%%%%%%%%%%%%%%

\subsection{The Extremal Reissner--Nordstr\"om Case}

%%%%%%%%%%%%%%%%%%%%%%%%%%%%%%%%%%%%%%%%%%%%%%%%%%%%%%%%%%%%%%%%%

Now we discuss the QNMs of the extremal Reissner--Nordstr\"om
$d$--dimensional black hole including first-order corrections. The
calculation for zeroth-order was done in \cite{natario-schiappa},
where they found that the $ q^2 \rightarrow m^2$ limit of RN QN
frequencies does not yield extremal RN QN frequencies. In this case,
the outer horizon approaches the inner horizon.

Again we define
$$
F(r_*)=e^{i \omega r_*}\Psi(r_*).
$$
Then the clockwise monodromy of $F(r_*)$ around the horizon $r=r_0$
can be obtained by continuing the coordinate $r$ analytically into
the complex plane, \textit{i.e.},
$$
\mathcal{M}(r_0)=e^{\frac{2\pi \omega} {k_0}},
$$
where $k_0= \frac{(d-3)^2}{2(d-2) m^{1/d}}$ .

Near the black hole singularity ($r \sim 0$), the tortoise
coordinate may be expanded as
$$
r_*=\int \frac{dr}{f(r)}=\frac{1}{2d-5}\frac{r^{2d-5}}{
m^2}+\frac{2}{3d-8}\frac{r^{3d-8}}{ m^3}+\cdots.
$$
Again, when we define $z=\omega r_*$, the potential for the three
different type perturbations can be expanded, respectively, as
(appendix \ref{appendixA})
\begin{equation}\label{potential-exrn}
V [ z ] \sim -\frac{\omega^2}{4z^2} \left \{1-j^2-W(j)
\left(\frac{z}{\omega}\right)^{(d-3)/(2d-5)}+\cdots \right\} ,
\end{equation}
where
\begin{eqnarray*}
W(j)=
 \begin{cases}
  W_{RNT}^{ex} &\text{$j=j_T$},\\
  W_{RNV^{\pm}}^{ex} &\text{$j=j_{V^{\pm}}$},\\
  W_{RNS^{\pm}}^{ex} &\text{$j=j_{S^{\pm}}$},
 \end{cases}
\end{eqnarray*}
 and the explicit expressions of $W_{RNT}^{ex}$, $W_{RNV^{\pm}}^{ex}$ and $W_{RNS^{\pm}}^{ex}$ can be found in appendix \ref{appendixA}.
Then the Schr\"{o}dinger-like wave equation (\ref{schrodinger}) with
the potential (\ref{potential-exrn}) can be depicted as
\begin{equation}\label{waveequationex}
 \left(\mathcal{H}_0
+\omega^{-\frac{d-3}{2d-5}}\mathcal{H}_1\right) \Psi=0,
\end{equation}
 where
$\mathcal{H}_0$ and $\mathcal{H}_1$ are defined as
\begin{eqnarray*}
\mathcal{H}_0=\frac{d^2}{dz^2}+\left[\frac{1-j^2}{4z^2}+1\right] , \
\ \ \mathcal{H}_1=-\frac{W(j)}{4}z^{-\frac{3d-7}{2d-5}}.
\end{eqnarray*}
\FIGURE[ht]{\label{StokesERN}
    \centering
    \psfrag{A}{$A$}
    \psfrag{B}{$B$}
    \psfrag{r0}{$r_0$}
    \psfrag{r1}{$r_1$}
    \psfrag{r2}{$r_2$}
    \psfrag{Re}{$\re$}
    \psfrag{Im}{$\im$}
    \psfrag{contour}{contour}
    \psfrag{Stokes line}{Stokes line}
    \epsfxsize=.6\textwidth
    \leavevmode
    \epsfbox{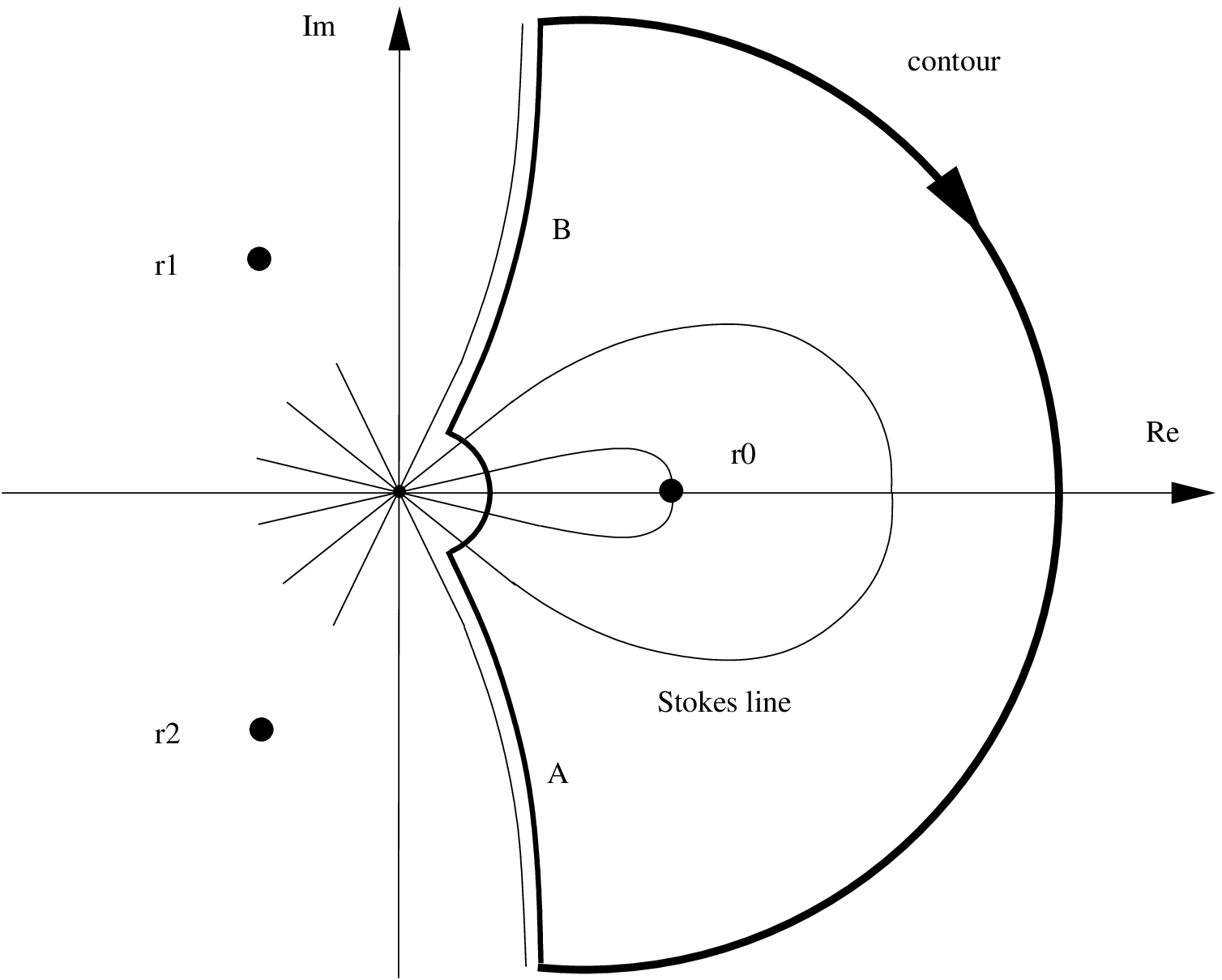}
\caption{Stokes line for the extremal Reissner--Nordstr\"om black
hole, along with the chosen contour for monodromy matching, in the
case $d=6$ ( we refer the reader to \cite{natario-schiappa} for
detail, and a more complete list of figures in dimensions $d=4$,
$d=5$, $d=6$ and $d=7$ ).} }
 Taking into account of $\omega\rightarrow\infty$, the zeroth-order wave
equation becomes
$$
\mathcal{H}_0 \Psi^{(0)}=0,
$$
with general solutions in the form of Eq. (\ref{solution}).
According to the boundary conditions (\ref{bounder}), one can define
$$
F^{(0)}(z)=F_+^{(0)}(z)-e^{-\pi j i/2}F_-^{(0)}(z),
$$
which approaches $-e^{-i\alpha_+}\sin \frac{j\pi}{2}$ as $z
\rightarrow +\infty$. This holds at point $A$ in
Figure\ref{StokesERN}. By going around an arc of angle of
$\frac{5\pi}{2d-5}$ in complex $r$--plane (rotating from $A$ to
$B$), $z$ rotates through an angle of $5\pi$ in $z$--plane, leading
to the wave function
\begin{equation*}
F^{(0)}(e^{5\pi i}z)=-e^{-i\alpha_+}\sin\frac{5j\pi}2.
\end{equation*}
Consequently, we deduce that
$$
e^{\frac{2\pi\omega}{k}}=\frac{\sin\frac{5j\pi}2}{\sin\frac{j\pi}2},
$$
leading to the generic expression of $d$-dimensional QN frequencies
$$
\frac{\omega}{T}=2n\pi i +\ln(1+2\cos j\pi+2\cos 2j\pi),
$$
where $T=2\pi k_0$.

Next, we calculate the first-order correction to the asymptotic
frequencies. Again we expand the wave function to the first order in
$1/\omega^{(d-3)/(2d-5)}$ as
$$
\Psi=\Psi^{(0)}+\frac 1 {\omega^{(d-3)/(2d-5)}} \Psi^{(1)}.
$$
Then one can rewrite Eq. (\ref{waveequationex}) as
\begin{equation*}
 \mathcal{H}_0
\Psi^{(1)}+\mathcal{H}_1 \Psi^{(0)}=0.
\end{equation*}
The general solution of this equation, as mentioned in the last
subsection, is
\begin{equation}\label{solutionern}
\Psi_{\pm}^{(1)}=\mathcal{C}\Psi_+^{(0)}\int_0^z
\Psi_-^{(0)}\mathcal{H}_1
\Psi_{\pm}^{(0)}-\mathcal{C}\Psi_-^{(0)}\int_0^z
\Psi_+^{(0)}\mathcal{H}_1 \Psi_{\pm}^{(0)}.
\end{equation}
The behavior as $z \gg 1$ is found to be
\begin{equation*}
\Psi_{\pm}^{(1)}(z) \sim c_{- \pm}\cos (z-\alpha_+)-c_{+ \pm}\cos
(z-\alpha_-).
\end{equation*}
After defining
\begin{eqnarray*}
\Psi &=& \Psi_+^{(0)} +\frac
1{\omega^{(d-3)/(2d-5)}}\Psi_+^{(1)}-\\
&&-e^{-ij\pi/2}\left(1-\frac{\xi}{\omega^{(d-3)/(2d-5)}}\right)\left(\Psi_-^{(0)}
+\frac 1{\omega^{(d-3)/(2d-5)}}\Psi_-^{(1)} \right),
\end{eqnarray*}
and using boundary condition (\ref{bounder}), one finds that the
wave function $F(z)$ approaches
$$
 F(z) \sim
-e^{-i \alpha_+} \sin \frac{j \pi}{2}
\left[1-\frac{\xi_-}{\omega^{(d-3)/(2d-5)}}\right],
$$
as $z \rightarrow \infty$.\\
 The function $\Psi_{\pm}^{(1)}$ defined
in (\ref{solutionern}) follows that
$$
\Psi_{\pm}^{(1)}=z^{1\pm j/2+\eta}G_{\pm}(z),
$$
where $\eta=-1/(4d-10)$, and $G_{\pm}(z)$ are even analytic
functions of $z$. By going around an arc of angle of
$\frac{5\pi}{2d-5}$ in complex $r$--plane, $z$ rotates through an
angle of $5\pi$ in $z$--plane, leading to the wave function
\begin{eqnarray*}
F(e^{5\pi i}z)&=& e^{-2iz}B+\frac12
\left(e^{9i\alpha_+}-e^{-ij\pi/2}e^{9i\alpha_-}\right)\\
&&+\frac{e^{5\eta \pi i}}{2\omega^{(d-3)/(2d-5)}}
\left[-i\xi_+e^{9i\alpha_+}-i\xi_-e^{-ij\pi/2}e^{9i\alpha_-}+\xi
e^{-5\eta \pi i}e^{-ij\pi/2}e^{9i\alpha_-}\right],
\end{eqnarray*}
where $B$ is a constant which is not needed in our calculation
because $e^{-2iz}\rightarrow 0$ as $z\rightarrow+\infty$. Finally,
we have
\begin{equation*}
e^{\frac{2\pi\omega}{k}}=
[1+2\cos(j\pi)+2\cos(2j\pi)]\left(1+\frac{\xi_--K}{\omega^{(d-3)/(2d-5)}}\right),
\end{equation*}
where
$$
K=e^{(17+10\eta)\pi i/4}\left[\xi_+\sin\left(\frac{3+10\eta}{4}
\pi\right)-\xi\cos\left(\frac{3+10\eta}{4}
\pi\right)\cot(\frac{5j\pi}2)+i\xi_-\cos\left(\frac{3+10\eta}{4}
\pi\right)\right],
$$
leading to the generic expression of $d$-dimensional QN
frequencies
$$
\frac{\omega}{T}=2n\pi i +\ln\left[1+2\cos (j\pi)+2\cos(2
j\pi)\right]+\frac{corr_d}{n^{(d-3)/(2d-5)}}.
$$
For $d=4$, one obtains
\begin{equation}\label{correxrn}
corr_4=\frac{\sqrt{3}i-1}{4\sqrt[3]{k_0}}
\left[\sqrt{3}\xi_+-\sqrt{3}\xi_--\xi\cot(5j\pi/2)\right].
\end{equation}
In order to obtain an explicit expression of $corr_4$, we need the
integral
$$\mathcal{J}(\nu, \mu)\equiv \int_0^{\infty}dz z^{-2/3}
J_{\nu}(z)J_{\mu}(z)=\frac{\Gamma(\frac23)\Gamma(\frac{
\mu+\nu+1/3}2)}{\sqrt[3] {4}\Gamma(\frac{\nu-\mu+5/3}2)\Gamma(\frac{
\mu+\nu+5/3}2)\Gamma(\frac{\mu-\nu+5/3}2)}.
$$
As a result, we obtain
\begin{eqnarray*}
\xi_+ &=& \frac{\pi W(j)}{8\sin(j\pi/2)}\left[\mathcal{J}(j/2,j/2)
e^{ij\pi/2} -
\mathcal{J}(j/2,-j/2)\right],\\
\xi_- &=& \frac{\pi W(j)}{8\sin(j\pi/2)}\left[\mathcal{J}(-j/2,-j/2)
e^{-ij\pi/2} - \mathcal{J}(j/2,-j/2)\right],
\end{eqnarray*}
Immediately, one has
$$
corr_4 = -\frac{W{(j)}}{8\sqrt[3]{4k_0}\pi^2}\times\frac{\cos
\frac{5j\pi}2-\cos
\frac{7j\pi}2-(1-\sqrt{3}i)\cos\frac{3j\pi}2}{\sin\frac{j\pi}2\sin\frac{5
j\pi}2}\cdot\Gamma(2/3)\Gamma(1/6+j/2) \Gamma(1/6-j/2)\Gamma^2(1/6).
$$
In this way, one can easily obtain the explicit expression of
$corr_4$ of Eq. (\ref{correxrn}).  For tensor and scalar type
perturbations, $W_{RNT}^{ex}=W_{RNS^{\pm}}^{ex}=0$, so $corr_4=0$,
leading to the same results as the zeroth-order asymptotic QN
frequencies. For vector type perturbations , we have
$j\rightarrow5/3$. Strangely, this also leads to $ corr_4
\rightarrow 0 $. It is interesting to investigate whether the
first-order corrections for any $d>3$ extremal RN black holes have a
same behavior---they all approach zero, independent on the
dimension. Another point is that in the limit of $ q^2\rightarrow
m^2$, $corr_4$ of RN black hole does not approaches zero---they have
different correction terms.

%%%%%%%%%%%%%%%%%%%%%%%%%%%%%%%%%%%%%%%%%%%%%%%%%%%%%%%%%%%%%%%%%

\subsection{The Schwarzschild de Sitter Case}

%%%%%%%%%%%%%%%%%%%%%%%%%%%%%%%%%%%%%%%%%%%%%%%%%%%%%%%%%%%%%%%%%

Now we compute the quasinormal modes of the Schwarzschild dS
$d$--dimensional black hole including first-order corrections. We
start with the zeroth-order calculation. For Schwarzschild de
Sitter(SdS) black hole, we have
$$
f(r)=1-\frac{2 m}{r^{d-3}}-\lambda r^2,
$$
with the roots
$$
r_n=r_H, r_C, r_1, r^*_1,\cdots, r_{\frac{d-4}2}, r^*_{\frac{d-4}2},
\tilde{r},
$$
where $\lambda >0$ is the black hole background parameter related to
the cosmological constant $\Lambda$ by
$$
\Lambda=\frac12 (d-1)(d-2)\lambda,
$$
and $r^*_n$ represents the conjugate of $r_n$. Here we have defined
$$
\tilde{r}=-\left(r_H+r_C+\sum_{i=1}^{(d-4)/2}(r_i+r^*_i)\right).
$$
The clockwise monodromy of $\Psi(r_*)$ around the event horizon
$r=r_H$ and the cosmological horizon $r=r_C$ can be obtained by
continuing the coordinate $r$ analytically into the complex plane,
respectively
\begin{eqnarray*}
\mathcal{M}(r_H) &=& e^{\frac{\pi \omega} {k_H}},\\
\mathcal{M}(r_C) &=& e^{\frac{-\pi \omega} {k_C}},
\end{eqnarray*}
where $k_H= \frac12 f'(r_H) $ and $k_C= \frac12 f'(r_C)$ are the
surface gravity at the event horizon and the cosmological horizon, respectively. \\
Near the black hole singularity ($r \sim 0$), the tortoise
coordinate may be expanded as
$$
r_*=\int \frac{dr}{f(r)}=-\frac{1}{d-2}\frac{r^{d-2}}{2
m}-\frac{1}{2d-5}\frac{r^{2d-5}}{(2 m)^2}+\cdots.
$$
After defining $z=\omega r_*$, the potential for the three different
type perturbations can be expanded, respectively, as (appendix
\ref{appendixA})
\begin{equation}\label{potential-sds}
V [ z ] \sim -\frac{\omega^2}{4z^2} \left \{1-j^2-W(j)
\left(\frac{z}{\omega}\right)^{(d-3)/(d-2)}+\cdots \right\} ,
\end{equation}
where
\begin{eqnarray*}
W(j)=
 \begin{cases}
  W_{SdST} &\text{$j=j_T$},\\
  W_{SdSV} &\text{$j=j_{V}$},\\
  W_{SdSS} &\text{$j=j_{S}$},
 \end{cases}
\end{eqnarray*}
 and the explicit expressions of $W_{SdST}$, $W_{SdSV}$ and $W_{SdSS}$ can be found in appendix \ref{appendixA}.
Then the Schr\"{o}dinger-like wave equation (\ref{schrodinger}) with
the potential (\ref{potential-sds}) can be depicted as
\begin{equation}\label{waveequationsds}
 \left(\mathcal{H}_0
+\omega^{-\frac{d-3}{d-2}}\mathcal{H}_1\right) \Psi=0,
\end{equation}
 where
$\mathcal{H}_0$ and $\mathcal{H}_1$ are defined as
\begin{eqnarray*}
\mathcal{H}_0=\frac{d^2}{dz^2}+\left[\frac{1-j^2}{4z^2}+1\right] , \
\ \ \mathcal{H}_1=-\frac{W(j)}{4}z^{-\frac{d-1}{d-2}}.
\end{eqnarray*}
Obviously, the zeroth-order wave equation can be written as
$$
\mathcal{H}_0 \Psi^{(0)}=0,
$$
with general solutions in the form of
$$
\Psi^{(0)}(z)=A_+\sqrt{\frac{\pi z}{2}}J_{j/2}(z)+A_-\sqrt{\frac{\pi
z}{2}}J_{-j/2}(z).
$$
As one lets $z\rightarrow +\infty$, the wave function approaches
\begin{equation*}
\Psi^{(0)}(z)\sim
\left(A_+e^{-i\alpha_+}+A_-e^{-i\alpha_-}\right)\frac{e^{iz}}2+
\left(A_+e^{i\alpha_+}+A_-e^{i\alpha_-}\right)\frac{e^{-iz}}2.
\end{equation*}

\FIGURE[ht]{\label{StokesSdS}
    \centering
    \psfrag{A}{$A$}
    \psfrag{B}{$B$}
    \psfrag{RH}{$r_H$}
    \psfrag{RC}{$r_C$}
    \psfrag{R}{$\tilde{r}$}
    \psfrag{g}{$r$}
    \psfrag{og}{$r^*$}
    \psfrag{Re}{$\re$}
    \psfrag{Im}{$\im$}
    \psfrag{contour}{contour}
    \psfrag{Stokes line}{Stokes line}
    \epsfxsize=.6\textwidth
    \leavevmode
    \epsfbox{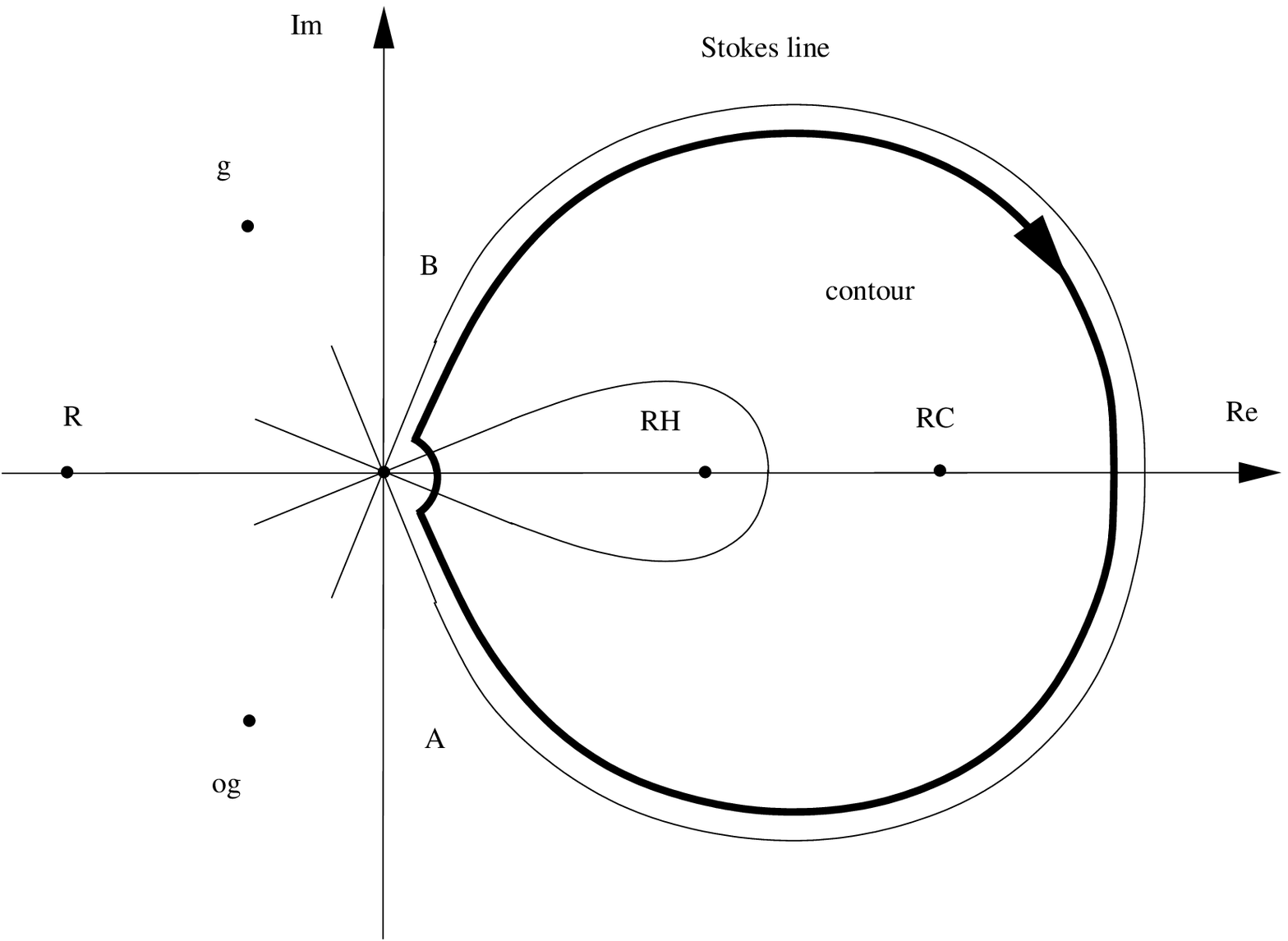}
\caption{Stokes line for the Schwarzschild de Sitter black hole,
along with the chosen contour for monodromy matching, in the case of
dimension $d=6$ ( we refer the reader to \cite{natario-schiappa} for
detail, and a more complete list of figures in dimensions $d=4$,
$d=5$, $d=6$ and $d=7$ ).} }
 This holds at point $A$ in Figure \ref{StokesSdS}. By going around an arc of angle of
$\frac{3\pi}{d-2}$ in complex $r$--plane (rotating from $A$ to $B$),
$z$ rotates through an angle of $3\pi$ in $z$--plane, leading to the
wave function
\begin{equation*}
\Psi^{(0)}(z) \sim
\left(A_+e^{7i\alpha_+}+A_-e^{7i\alpha_-}\right)\frac{e^{iz}}2+
\left(A_+e^{5i\alpha_+}+A_-e^{5i\alpha_-}\right)\frac{e^{-iz}}2.
\end{equation*}
Consequently, we have
\begin{eqnarray}
\frac{A_+ e^{7i\alpha_+} + A_- e^{7i\alpha_-}}{A_+ e^{-i\alpha_+} +
A_-
e^{-i\alpha_-}} e^{\frac{\pi\omega}{k_H}+\frac{\pi\omega}{k_C}}& =& \mathcal{M}(r_H)\mathcal{M}(r_C),\label{7sds0}\\
\frac{A_+ e^{5i\alpha_+} + A_- e^{5i\alpha_-}}{A_+ e^{i\alpha_+} +
A_- e^{i\alpha_-}} e^{-\frac{\pi\omega}{k_H}-\frac{\pi\omega}{k_C}}&
=& \mathcal{M}(r_H)\mathcal{M}(r_C).\label{5sds0}
\end{eqnarray}
Taking the condition for these equations to have nontrivial
solutions $(A_+,A_-)$ into account, one may easily obtain the final
results as
\begin{equation}\label{sdsresult1}
\cosh\left(\frac{\pi\omega}{k_C}-\frac{\pi\omega}{k_H}\right)+(1+2\cos
j\pi)
\cosh\left(\frac{\pi\omega}{k_H}+\frac{\pi\omega}{k_C}\right)=0.
\end{equation}
As discussed in Ref. \cite{natario-schiappa}, the Stokes lines for
$d=4$ and $d=5$ are a bit different from the case discussed in the
previous calculation for $d=6$. On account of the behavior of $r\sim
\infty$, they found that the final result for $d=4$ is the same as
Eq.(\ref{sdsresult1}). However, the result for $d=5$ is changed. For
$r\sim\infty$, the coefficient of $e^{-iz}$ in the formula of the
wave function keeps unchanged, while the coefficient of $e^{iz}$ in
there reverses sign as one rotates from the branch containing point
$B$ to the branch containing point $A$ in the contour. Therefore one
can complete this calculation by reversing the sign of Eq.
(\ref{7sds0}). In the end, we obtain
\begin{equation*}
\sinh\left(\frac{\pi\omega}{k_C}-\frac{\pi\omega}{k_H}\right)+(1+2\cos
j\pi)
\sinh\left(\frac{\pi\omega}{k_H}+\frac{\pi\omega}{k_C}\right)=0.
\end{equation*}

Next, we calculate the first-order correction to the asymptotic
frequencies. Again we expand the wave function to the first order in
$1/\omega^{(d-3)/(d-2)}$ as
$$
\Psi=\Psi^{(0)}+\frac 1 {\omega^{(d-3)/(d-2)}} \Psi^{(1)}.
$$
Then one can rewrite Eq. (\ref{waveequationsds}) as
\begin{equation*}
 \mathcal{H}_0
\Psi^{(1)}+\mathcal{H}_1 \Psi^{(0)}=0.
\end{equation*}
The general solution of this equation is
\begin{equation}\label{solutionsds}
\Psi_{\pm}^{(1)}=\mathcal{C}\Psi_+^{(0)}\int_0^z
\Psi_-^{(0)}\mathcal{H}_1
\Psi_{\pm}^{(0)}-\mathcal{C}\Psi_-^{(0)}\int_0^z
\Psi_+^{(0)}\mathcal{H}_1 \Psi_{\pm}^{(0)}.
\end{equation}
The behavior as $z \gg 1$, as mentioned above, is found to be
\begin{equation*}
\Psi_{\pm}^{(1)}(z) \sim c_{- \pm}\cos (z-\alpha_+)-c_{+ \pm}\cos
(z-\alpha_-).
\end{equation*}
After defining
$$
a_{\pm\pm}=c_{\pm\pm}\omega^{-\frac{d-3}{d-2}},\ \ \ \ u_{\pm}=1\pm
a_{+-},
$$
one finds that the wave function $\Psi(z)$ approaches
\begin{eqnarray*}
 \Psi(z)&=& \Psi^{(0)}(z)+ \omega^{-\frac{d-3}{d-2}}\Psi^{(1)}(z)\\
& \sim &
(A_+u_++A_-a_{--})\cos(z-\alpha_+)+(A_-u_--A_+a_{++})\cos(z-\alpha_-)\\
&=&
\left[(A_+u_++A_-a_{--})e^{-i\alpha_+}+(A_-u_--A_+a_{++})e^{-i\alpha_-}\right]\frac{e^{iz}}2+\\&
&\left[(A_+u_++A_-a_{--})e^{i\alpha_+}+(A_-u_--A_+a_{++})e^{i\alpha_-}\right]\frac{e^{-iz}}2
\end{eqnarray*}
as $z \rightarrow \infty$.\\
 The function $\Psi_{\pm}^{(1)}$ defined
in (\ref{solutionsds}) follows that
$$
\Psi_{\pm}^{(1)}=z^{1/2\pm j/2+\eta}G_{\pm}(z),
$$
where $\eta=(d-3)/(d-2)$, and $G_{\pm}(z)$ are even analytic
functions of $z$. By going around an arc of angle of
$\frac{3\pi}{d-2}$ in complex $r$--plane, $z$ rotates through an
angle of $3\pi$ in $z$--plane, leading to the wave function
\begin{eqnarray*}
\Psi(z) &\sim &
\left[(A_+v_++A_-a_{--}e^{-3ij\pi})e^{7i\alpha_+}+(A_-v_--A_+a_{++}e^{3ij\pi})e^{7i\alpha_-}\right]\frac{e^{3\eta\pi
i}e^{iz}}2+\\&&
\left[(A_+v_++A_-a_{--}e^{-3ij\pi})e^{5i\alpha_+}+(A_-v_--A_+a_{++}e^{3ij\pi})e^{5i\alpha_-}\right]\frac{e^{3\eta\pi
i}e^{-iz}}2.
\end{eqnarray*}
as $z \rightarrow -\infty$. Here we have defined
$v_{\pm}=e^{-3\eta\pi i}\pm a_{+-}$. We hence have the similar
formulae as shown in (\ref{7sds0}) and (\ref{5sds0}), \textit{i.e.},
\begin{eqnarray}
\frac{(A_+v_++A_-a_{--}e^{-3ij\pi})e^{7i\alpha_+}+(A_-v_--A_+a_{++}e^{3ij\pi})e^{7i\alpha_-}}
{(A_+u_++A_-a_{--})e^{-i\alpha_+}+(A_-u_--A_+a_{++})e^{-i\alpha_-}}
e^{\frac{\pi\omega}{k_H}+\frac{\pi\omega}{k_C}}& =& e^{-3\eta\pi
i}\mathcal{M}(r_H)\mathcal{M}(r_C),\label{7sds}\\
\frac{(A_+v_++A_-a_{--}e^{-3ij\pi})e^{5i\alpha_+}+(A_-v_--A_+a_{++}e^{3ij\pi})e^{5i\alpha_-}}
{(A_+u_++A_-a_{--})e^{i\alpha_+}+(A_-u_--A_+a_{++})e^{i\alpha_-}}
e^{-\frac{\pi\omega}{k_H}-\frac{\pi\omega}{k_C}}& =& e^{-3\eta\pi i}
\mathcal{M}(r_H)\mathcal{M}(r_C).\label{5sds1st}
\end{eqnarray}
In this way, one can easily obtain a set of equations with regard to
$A_+$ and $A_-$
\begin{eqnarray*}
\left(s_1e^{7i\alpha_+}-s_2e^{-\frac{2\pi\omega}{k_C}}e^{-i\alpha_+}\right)A_++
\left(s_3e^{7i\alpha_-}-s_4e^{-\frac{2\pi\omega}{k_C}}e^{-i\alpha_-}\right)A_- &=& 0,\\
\left(s_5e^{5i\alpha_+}-s_6e^{\frac{2\pi\omega}{k_H}}e^{i\alpha_+}\right)A_++
\left(s_7e^{5i\alpha_-}-s_8e^{\frac{2\pi\omega}{k_H}}e^{i\alpha_-}\right)A_-
&=& 0,
\end{eqnarray*}
where
\begin{eqnarray*}
s_1 &=& (v_+-a_{++}e^{-\frac{ij\pi}2})e^{3\eta\pi i},\ \ \ s_2=(u_+-a_{++}e^{\frac{ij\pi}2}),\\
s_3 &=& (v_-+a_{--}e^{\frac{ij\pi}2})e^{3\eta\pi i},\ \ \ s_4=(u_-+a_{--}e^{-\frac{ij\pi}2}),\\
s_5 &=& (v_+-a_{++}e^{\frac{ij\pi}2})e^{3\eta\pi i},\ \ \ s_6=(u_+-a_{++}e^{-\frac{ij\pi}2}),\\
s_7 &=& (v_-+a_{--}e^{-\frac{ij\pi}2})e^{3\eta\pi i},\ \ \ s_8=(u_-+a_{--}e^{\frac{ij\pi}2}).\\
\end{eqnarray*}
The condition for these equations to have nontrivial solutions
$(A_+,A_-)$ is then
\begin{align*}
\left|
\begin{array}{ccc}
s_1e^{7i\alpha_+}-s_2e^{-\frac{2\pi\omega}{k_C}}e^{-i\alpha_+} & \,\,\,\, & s_3e^{7i\alpha_-}-s_4e^{-\frac{2\pi\omega}{k_C}}e^{-i\alpha_-} \\
s_5e^{5i\alpha_+}-s_6e^{\frac{2\pi\omega}{k_H}}e^{i\alpha_+}   &  &
s_7e^{5i\alpha_-}-s_8e^{\frac{2\pi\omega}{k_H}}e^{i\alpha_-}
\end{array}
\right| = 0.
\end{align*}
After some algebra, we obtain
\begin{equation}\label{sds1st}
\mathcal{A}_1\cosh\left(\frac{\pi\omega}{k_C}-\frac{\pi\omega}{k_H}+\Delta_1\right)
+\mathcal{A}_2\cosh\left(\frac{\pi\omega}{k_H}+\frac{\pi\omega}{k_C}+\Delta_2\right)=0,
\end{equation}
where
\begin{eqnarray*}
\mathcal{A}_1 &=&
\sqrt{\left(s_1s_7e^{\frac{ij\pi}2}-s_3s_5e^{-\frac{ij\pi}2}\right)\left(s_4s_6e^{\frac{ij\pi}2}-s_2s_8e^{-\frac{ij\pi}2}\right)},\\
\mathcal{A}_2 &=&
\sqrt{\left(s_1s_8e^{\frac{3ij\pi}2}-s_3s_6e^{-\frac{3ij\pi}2}\right)\left(s_4s_5e^{\frac{3ij\pi}2}-s_2s_7e^{-\frac{3ij\pi}2}\right)},\\
\Delta_1 &=& \frac12
\left[\log\left(s_1s_7e^{\frac{ij\pi}2}-s_3s_5e^{-\frac{ij\pi}2}\right)-\log\left(s_4s_6e^{\frac{ij\pi}2}-s_2s_8e^{-\frac{ij\pi}2}\right)\right],\\
\Delta_2 &=& \frac12
\left[\log\left(s_1s_8e^{\frac{3ij\pi}2}-s_3s_6e^{-\frac{3ij\pi}2}\right)-\log\left(s_4s_5e^{\frac{3ij\pi}2}-s_2s_7e^{-\frac{3ij\pi}2}\right)\right].
\end{eqnarray*}
For zeroth-order asymptotic QN frequencies, one has $a_{\pm\pm}=0$,
and hence $u_{\pm}=1$ and $v_{\pm}=e^{-3\eta\pi i}$. As a result,
formula (\ref{sds1st}) reduces to (\ref{sdsresult1}).

As mentioned formerly, the Stokes lines for $d=4$ and $d=5$ are a
bit different from the case discussed in the previous calculation
for $d=6$. Resorting to the behavior of $r \sim \infty$, one finds
that the final result for $d=4$ is shown to be the same as
Eq.(\ref{sds1st}). However, the result for $d=5$ is changed. For
$r\sim\infty$, the coefficient of $e^{-iz}$ in the formula of the
wave function keeps unchanged, while the coefficient of $e^{iz}$ in
there reverses sign as one rotates from the branch containing point
$B$ to the branch containing point $A$ in the contour. Therefore one
can complete this calculation by reversing the sign of Eq.
(\ref{7sds}). In the end, we obtain
\begin{equation*}
\mathcal{A}_1\sinh\left(\frac{\pi\omega}{k_C}-\frac{\pi\omega}{k_H}+\Delta_1\right)
+\mathcal{A}_2\sinh\left(\frac{\pi\omega}{k_H}+\frac{\pi\omega}{k_C}+\Delta_2\right)=0.
\end{equation*}

For $d=4$ SdS black holes, one has $r_C \gg r_H$ as $0<\lambda \ll
1$, as a result, one has $k_H\gg k_C$. In this case, we can neglect
$k_C$ compared to $k_H$.  Taking Eq. (\ref{sdsresult1}) into
account, one finds the asymptotic QN frequencies in this case is the
solutions of this formula
$$
e^{\frac{\pi\omega}{k_C}}+e^{-\frac{\pi\omega}{k_C}}=0.
$$
So, we have $ \frac{\omega}{T_C}=(2n+1)\pi i$. However, as one lets
$\lambda$ approaches its extremal value, \textit{i.e.},
$\lambda\rightarrow 1/27$ (with $ m=1$), one has $k_H \approx k_C$.
So, in this limit, we can obtain the asymptotic QN frequencies by
solving this formula
$$
e^{2\frac{\pi\omega}{k_C}}+e^{-2\frac{\pi\omega}{k_C}}=0,
$$
which in turn derives $ \frac{\omega}{T_C}=\frac12(2n+1)\pi i$,
where $T_C$ is the Hawking temperature at cosmological horizon. This
reminds us to conjecture that the frequencies have values between
these two extremal values, \textit{i.e.},
\begin{equation}\label{chi1}
\frac{\omega}{T_C}\simeq \chi(2n+1)\pi i+{\mathbb{R}}{\mathrm{e}}\,
\omega,\ \ \ \ \ \frac12<\chi<1,
\end{equation}
where $\chi$ is a parameter closely related to the cosmological
constant $\lambda$. In fact, it would be an interesting work to
investigate the relationship between these two parameters, by
analytical methods or numerical ones. ${\mathbb{R}}{\mathrm{e}}\,
\omega$ is the real part of the frequency.

In order to obtain the explicit expressions of $\mathcal{A}_1$,
$\mathcal{A}_2$, $\Delta_1$ and $\Delta_2$, we need the integral
$$\int_0^{\infty}dz z^{-1/2}
J_{\nu}(z)J_{\mu}(z)=\frac{\sqrt{\pi/2}\Gamma(\frac{
\mu+\nu+1/2}2)}{\Gamma(\frac{\nu-\mu+3/2}2)\Gamma(\frac{
\mu+\nu+3/2}2)\Gamma(\frac{\mu-\nu+3/2}2)}.
$$
As a result, we obtain
\begin{eqnarray*}
s_1 &=& 1-2iU(j)\left(\cos\frac{j\pi}2-\sin\frac{j\pi}2\right),\ \ \
\ \
s_2 = 1-2iU(j)\left(\cos\frac{j\pi}2-\sin\frac{j\pi}2\right),\\
s_3 &=& 1-2iU(j)\left(\cos\frac{j\pi}2+\sin\frac{j\pi}2\right),\ \ \
\ \
s_4 = 1-2iU(j)\left(\cos\frac{j\pi}2+\sin\frac{j\pi}2\right),\\
s_5 &=& 1-2U(j)\left(\cos\frac{j\pi}2-\sin\frac{j\pi}2\right),\ \ \
\ \
s_6 = 1+2U(j)\left(\cos\frac{j\pi}2-\sin\frac{j\pi}2\right),\\
s_7 &=& 1-2U(j)\left(\cos\frac{j\pi}2+\sin\frac{j\pi}2\right),\ \ \
\ \ s_8 = 1+2U(j)\left(\cos\frac{j\pi}2+\sin\frac{j\pi}2\right),
\end{eqnarray*}
where
$$
U(j)=\frac{W(j)}{32\pi^{3/2}\sqrt{(2n+1)\chi k_C}}
\Gamma^2(1/4)\Gamma(1/4+j/2)\Gamma(1/4-j/2).
$$
After some algebra, we obtain the explicit expressions of
$\mathcal{A}_1$, $\mathcal{A}_2$, $\Delta_1$ and $\Delta_2$ in terms
of $o(1/\sqrt{n})$
\begin{eqnarray*}
\mathcal{A}_1 &\simeq& 2i\sin\frac{j\pi}2,\\
\mathcal{A}_2 &\simeq& 2i\sin\frac{3j\pi}2\left[1-2iU(j)\left(\cos
j\pi/2+\frac{\cos3j\pi/2}{1+2\cos
j\pi}\right)\right],\\
\Delta_1 &\simeq& 0,\\
\Delta_2 &\simeq& 2U(j)\left(\cos j\pi/2+\frac{\cos3j\pi/2}{1+2\cos
j\pi}\right).
\end{eqnarray*}
All type perturbations, including tensor and scalar type
perturbations ($j\rightarrow 0^+$), and vector type perturbation
($j\rightarrow 2$) have a same behavior that $\mathcal{A}_1$ and
$\mathcal{A}_2$ approach zero in the limit. This makes Eq.
(\ref{sds1st}) automatically satisfied. As shown in
\cite{natario-schiappa}, we first regard $j$ as nonzero, then take
the limit as $j\rightarrow 0$. In the end, Eq. (\ref{sds1st})
becomes
\begin{equation}\label{sds1st4}
\cosh\left(\frac{\pi\omega}{k_C}-\frac{\pi\omega}{k_H}\right)
+(3-8iU(j))\cosh\left(\frac{\pi\omega}{k_H}+\frac{\pi\omega}{k_C}+\frac{8U(j)}3\right)=0.
\end{equation}
Inserting the expressions of $W(j)$ (see Appendix \ref{appendixA}),
we can easily obtain the explicit expression of $U(j)$. In this way,
the first-order correction to asymptotic QN frequencies of SdS black
holes can be obtained by evaluating Eq. (\ref{sds1st4}). The result
show that the correction term is closely related to $\ell$,
$\lambda$ (though the parameter $\chi$ and $k_C$), and of course,
$n$.

Numerical calculation on this case for very highly damped overtone
first appeared in \cite{yoshida-futamase} and later in
\cite{konoplya-zhidenko}. An analytical formula for four dimensional
case was deduced in \cite{cns}. They found that the analytical
results are in good agreement with the numerical results.  However,
it is interesting to perform further checks to our first-order
corrected results both in four and higher dimensional black hole
spacetimes.

%%%%%%%%%%%%%%%%%%%%%%%%%%%%%%%%%%%%%%%%%%%%%%%%%%%%%%%%%%%%%%%%%

\subsection{The Reissner--Nordstr\"om de Sitter Case}

%%%%%%%%%%%%%%%%%%%%%%%%%%%%%%%%%%%%%%%%%%%%%%%%%%%%%%%%%%%%%%%%%

Now we compute the quasinormal modes of the RN dS $d$--dimensional
black hole including first-order corrections. We start with the
zeroth-order calculation. For RN de Sitter black hole, we have
$$
f(r)=1-\frac{2 m}{r^{d-3}}+\frac{ q^2}{r^{2d-6}}-\lambda r^2,
$$
with the roots
$$
r_n=r_+, r_-, r_C, r_1, r^*_1,\cdots, r_{d-4}, r^*_{d-4}, \tilde{r},
$$
where $\lambda (>0)$ is the black hole background parameter related
to the cosmological constant $\Lambda$ by
$$
\Lambda=\frac12 (d-1)(d-2)\lambda,
$$
and $r^*_n$ represents the conjugate of $r_n$. Here we have defined
$$
\tilde{r}=-\left(r_++r_-+r_C+\sum_{i=1}^{d-4}(r_i+r^*_i)\right).
$$
 The clockwise monodromy of $\Psi(r_*)$ around
the outer horizon $r=r_+$ and the cosmological horizon $r=r_C$ can
be obtained by continuing the coordinate $r$ analytically into the
complex plane, respectively
\begin{eqnarray*}
\mathcal{M}(r_+) &=& e^{\frac{\pi \omega} {k_+}},\\
\mathcal{M}(r_C) &=& e^{\frac{-\pi \omega} {k_C}},
\end{eqnarray*}
where $k_+= \frac12 f'(r_+) $ and $k_C= \frac12 f'(r_C)$ are the
surface gravity at the outer horizon and the cosmological horizon, respectively. \\
Near the black hole singularity ($r \sim 0$), the tortoise
coordinate may be expanded as
$$
r_*=\int \frac{dr}{f(r)}=\frac{1}{2d-5}\frac{r^{2d-5}}{ q^2}+\frac{2
m}{3d-8}\frac{r^{3d-8}}{ q^4}+\cdots,
$$
In this procedure, we have assumed $\frac r{r_0^-}\ll 1$, where
$r_0^-=\left( m-\sqrt{ m^2- q^2}\right)^{\frac1{d-3}}$, represents
the inner horizon of the RN black hole. Again we must expand the
potential to the first order in
$1/\left[(r_0^-)^{2d-5}\omega\right]^{(d-3)/(2d-5)}$ instead of
$1/\omega^{(d-3)/(2d-5)}$. After defining $z=\omega r_*$, the
potential for the three different type perturbations can be
expanded, respectively, as (appendix \ref{appendixA})
\begin{equation}\label{potential-rnds}
V [ z ] \sim -\frac{\omega^2}{4z^2} \left \{1-j^2-W(j)
\left(\frac{z}{(r_0^-)^{2d-5}\omega}\right)^{(d-3)/(2d-5)}+\cdots
\right\} ,
\end{equation}
where
\begin{eqnarray*}
W(j)=
 \begin{cases}
  W_{RNdST} &\text{$j=j_T$},\\
  W_{RNdSV^{\pm}} &\text{$j=j_{V^{\pm}}$},\\
  W_{RNdSS^{\pm}} &\text{$j=j_{S^{\pm}}$},
 \end{cases}
\end{eqnarray*}
 and the explicit expressions of $W_{RNdST}$, $W_{RNdSV^{\pm}}$ and $W_{RNdSS^{\pm}}$ can be found in appendix \ref{appendixA}.
Then the Schr\"{o}dinger-like wave equation (\ref{schrodinger}) with
the potential (\ref{potential-rnds}) can be depicted as
\begin{equation}\label{waveequationrnds}
 \left(\mathcal{H}_0
+\left[(r_0^-)^{2d-5}\omega\right]^{-\frac{d-3}{2d-5}}\mathcal{H}_1\right)
\Psi=0,
\end{equation}
 where
$\mathcal{H}_0$ and $\mathcal{H}_1$ are defined as
\begin{eqnarray*}
\mathcal{H}_0=\frac{d^2}{dz^2}+\left[\frac{1-j^2}{4z^2}+1\right] , \
\ \ \mathcal{H}_1=-\frac{W(j)}{4}z^{-\frac{3d-7}{2d-5}}.
\end{eqnarray*}
Obviously, the zeroth-order wave equation can be written as
$$
\mathcal{H}_0 \Psi^{(0)}=0,
$$
with general solutions in the form of
$$
\Psi^{(0)}(z)=A_+\sqrt{\frac{\pi z}{2}}J_{j/2}(z)+A_-\sqrt{\frac{\pi
z}{2}}J_{-j/2}(z).
$$
As one lets $z\rightarrow +\infty$, the wave function approaches
\begin{equation}\label{rnds1}
\Psi^{(0)}(z)\sim
\left(A_+e^{-i\alpha_+}+A_-e^{-i\alpha_-}\right)\frac{e^{iz}}2+
\left(A_+e^{i\alpha_+}+A_-e^{i\alpha_-}\right)\frac{e^{-iz}}2.
\end{equation}

\FIGURE[ht]{\label{StokesRNdS}
    \centering
    \psfrag{A}{$A$}
    \psfrag{B}{$B$}
    \psfrag{r0}{$r_-$}
    \psfrag{R0}{$r_+$}
    \psfrag{RC}{$r_C$}
    \psfrag{R}{$\tilde{r}$}
    \psfrag{r1}{$r_1$}
    \psfrag{r2}{$r^*_1$}
    \psfrag{R1}{$r_2$}
    \psfrag{R2}{$r^*_2$}
    \psfrag{Re}{$\re$}
    \psfrag{Im}{$\im$}
    \psfrag{contour}{contour}
    \psfrag{Stokes line}{Stokes line}
    \epsfxsize=.6\textwidth
    \leavevmode
    \epsfbox{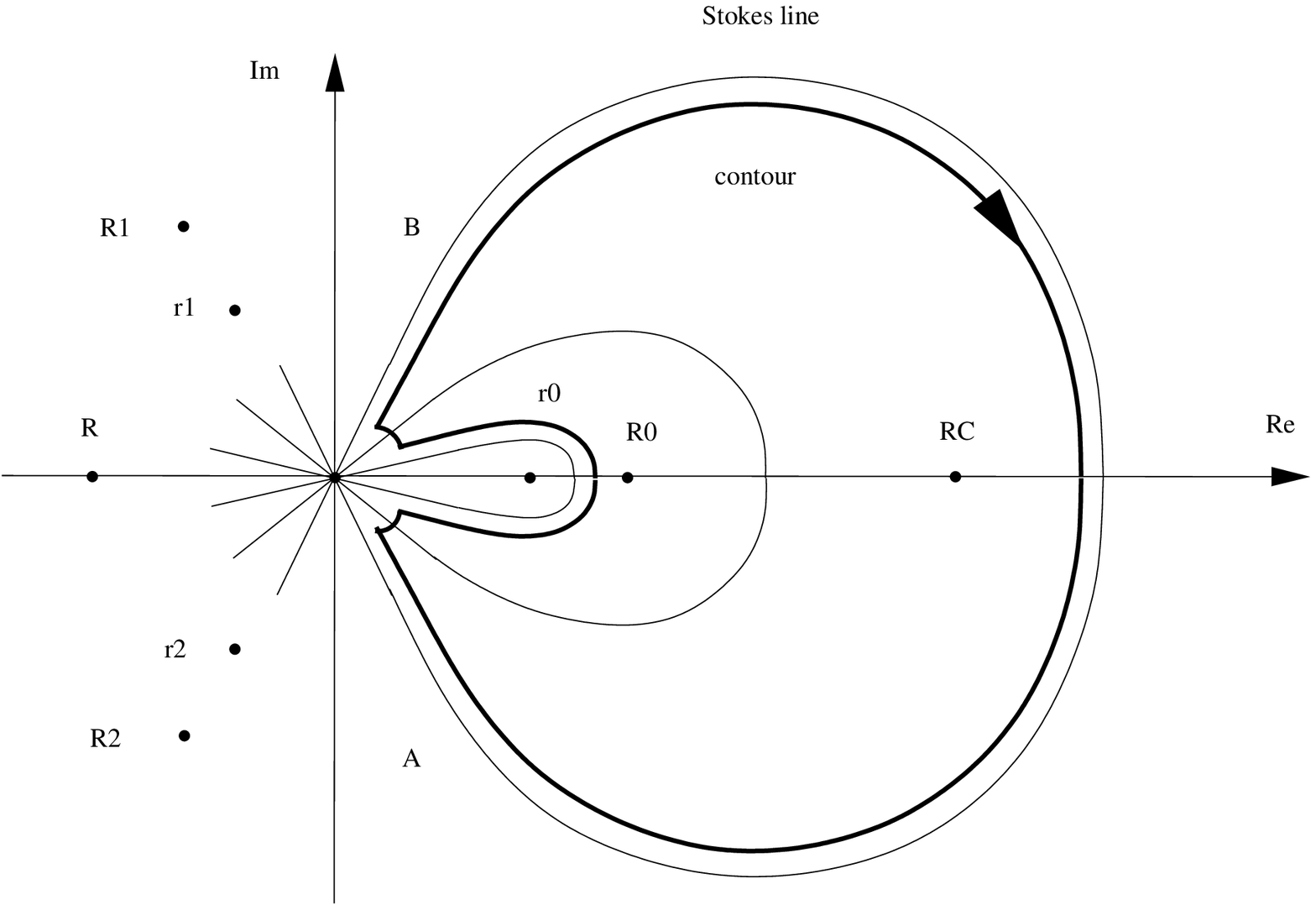}
\caption{Stokes line for the Reissner--Nordstr\"om de Sitter black
hole, along with the chosen contour for monodromy matching, in the
case of dimension $d=6$ ( we refer the reader to
\cite{natario-schiappa} for detail, and a more complete list of
figures in dimensions $d=4$, $d=5$, $d=6$ and $d=7$ ).} }

 This holds at point $A$ in Figure \ref{StokesRNdS}. By going around an arc of angle of
$\frac{2\pi}{2d-5}$ in complex $r$--plane (rotating from $A$ to the
next branch), $z$ rotates through an angle of $2\pi$ in $z$--plane,
leading to the wave function
\begin{equation*}
\Psi^{(0)}(z) \sim
\left(A_+e^{3i\alpha_+}+A_-e^{3i\alpha_-}\right)\frac{e^{iz}}2+
\left(A_+e^{5i\alpha_+}+A_-e^{5i\alpha_-}\right)\frac{e^{-iz}}2.
\end{equation*}
As one follows the contour around the inner horizon $r=r_-$, the
wave function will be of the form
\begin{equation*}
\Psi^{(0)}(z-\delta)=B_+\sqrt{\frac{\pi(z-\delta)}2}J_{
j/2}(z-\delta)+B_-\sqrt{\frac{\pi(z-\delta)}2}J_{-j/2}(z-\delta),
\end{equation*}
where
$$
\delta = \frac{2\omega\pi i}{f'(r_-)} = \frac{\omega\pi i}{k_-},
$$
and $ k_- = \frac12 f'(r_-) $ is the surface gravity at the inner
horizon. Notice that $z-\delta< 0$ on this branch, one can easily
obtain
\begin{equation}\label{rnds2}
\Psi^{(0)}(z-\delta)\sim\left(B_+e^{i\alpha_+}+B_-e^{i\alpha_-}\right)\frac{e^{i(z-\delta)}}2+
\left(B_+e^{-i\alpha_+}+B_-e^{-i\alpha_-}\right)\frac{e^{-i(z-\delta)}}2,
\end{equation}
where we have used
\begin{equation*}
J_\nu(z)=\sqrt{\frac{2}{\pi z}} \cos\left(z+\frac{\nu
\pi}{2}+\frac{\pi}4 \right) , \qquad z \ll -1.
\end{equation*}
Since on the same branch, we must let
$\Psi^{(0)}(z-\delta)=\Psi^{(0)}(e^{2\pi i}z)$, and hence it is
easily seen from Eqs.(\ref{rnds1}) and (\ref{rnds2}) that we have
\begin{eqnarray}
&& A_+e^{3i\alpha_+} + A_-e^{3i\alpha_-} = (B_+  e^{i\alpha_+} + B_- e^{i\alpha_-}) e^{-i\delta},\label{BB}\\
&& A_+e^{5i\alpha_+} + A_-e^{5i\alpha_-} = (B_+  e^{-i\alpha_+} +
B_- e^{-i\alpha_-}) e^{i\delta}.
\end{eqnarray}
Finally, we should rotate to the branch containing point $B$. This
makes $z - \delta$ rotate through an angle of $2\pi$, leading to the
wave function
\begin{equation*}
\Psi^{(0)}(z-\delta)\sim\left(B_+e^{5i\alpha_+}+B_-e^{5i\alpha_-}\right)\frac{e^{i(z-\delta)}}2+
\left(B_+e^{3i\alpha_+}+B_-e^{3i\alpha_-}\right)\frac{e^{-i(z-\delta)}}2.
\end{equation*}
Consequently, we have
\begin{eqnarray}
\frac{B_+ e^{5i\alpha_+} + B_- e^{5i\alpha_-}}{A_+ e^{-i\alpha_+} +
A_-
e^{-i\alpha_-}}e^{-i\delta} e^{\frac{\pi\omega}{k_+}+\frac{\pi\omega}{k_C}}& =& \mathcal{M}(r_+)\mathcal{M}(r_C),\label{7rnds0}\\
\frac{B_+ e^{3i\alpha_+} + B_- e^{3i\alpha_-}}{A_+ e^{i\alpha_+} +
A_- e^{i\alpha_-}}e^{i\delta}
e^{-\frac{\pi\omega}{k_H}-\frac{\pi\omega}{k_C}}& =&
\mathcal{M}(r_+)\mathcal{M}(r_C).\label{AB}
\end{eqnarray}
Taking the condition for these equations (\ref{BB}$\sim$ \ref{AB})
to have nontrivial solutions $(A_+,A_-,B_+,B_-)$ into account, one
can easily obtain the final results
\begin{equation}\label{rndsresult1}
\cosh\left(\frac{\pi\omega}{k_+}-\frac{\pi\omega}{k_C}\right)+\left(1+2\cos(j\pi)\right)
\cosh\left(\frac{\pi\omega}{k_+}+\frac{\pi\omega}{k_C}\right)+2(1+\cos(j\pi))
\cosh\left(\frac{\pi\omega}{k_+}+\frac{2\pi\omega}{k_-}+\frac{\pi\omega}{k_C}\right)=0.
\end{equation}

As discussed in ref \cite{natario-schiappa}, the Stokes lines for
$d=4$ and $d=5$ are a bit different from the case discussed in the
previous calculation for $d=6$. On account of the behavior of $r\sim
\infty$, they found that the final result for $d=4$ is the same as
Eq.(\ref{rndsresult1}). However, the result for $d=5$ is changed.
For $r\sim\infty$, the coefficient of $e^{-iz}$ in the formula of
the wave function keeps unchanged, while the coefficient of $e^{iz}$
in there reverses sign as one rotates from the branch containing
point $B$ to the branch containing point $A$ in the contour.
Therefore one can complete this calculation by reversing the sign of
Eq. (\ref{7rnds0}). In the end, we obtain
\begin{equation*}
\sinh\left(\frac{\pi\omega}{k_C}-\frac{\pi\omega}{k_+}\right)+\left(1+2\cos(2\pi/5)\right)
\sinh\left(\frac{\pi\omega}{k_+}+\frac{\pi\omega}{k_C}\right)+2(1+\cos(2\pi/5))
\sinh\left(\frac{\pi\omega}{k_+}+\frac{2\pi\omega}{k_-}+\frac{\pi\omega}{k_C}\right)=0.
\end{equation*}

Next, we calculate the first-order correction to the asymptotic
frequencies. Again we expand the wave function to the first order in
$1/\left[(r_0^-)^{2d-5}\omega\right]^{(d-3)/(2d-5)}$ as
$$
\Psi=\Psi^{(0)}+\frac 1
{\left[(r_0^-)^{2d-5}\omega\right]^{(d-3)/(2d-5)}} \Psi^{(1)}.
$$
Then one can rewrite Eq. (\ref{waveequationrnds}) as
\begin{equation}
 \mathcal{H}_0
\Psi^{(1)}+\mathcal{H}_1 \Psi^{(0)}=0.
\end{equation}
The general solution of this equation is
\begin{equation}\label{solutionrnds}
\Psi_{\pm}^{(1)}=\mathcal{C}\Psi_+^{(0)}\int_0^z
\Psi_-^{(0)}\mathcal{H}_1
\Psi_{\pm}^{(0)}-\mathcal{C}\Psi_-^{(0)}\int_0^z
\Psi_+^{(0)}\mathcal{H}_1 \Psi_{\pm}^{(0)}.
\end{equation}
The behavior as $z \gg 1$, as mentioned above, is found to be
\begin{equation*}
\Psi_{\pm}^{(1)}(z) \sim c_{- \pm}\cos (z-\alpha_+)-c_{+ \pm}\cos
(z-\alpha_-).
\end{equation*}
After defining
$$
a_{\pm\pm}=c_{\pm\pm}\left[(r_0^-)^{2d-5}\omega\right]^{-\frac{d-3}{2d-5}},\
\ \ \ u_{\pm}=1\pm a_{+-},
$$
one finds that the wave function $\Psi(z)$ approaches
\begin{eqnarray*}
 \Psi(z)&=& \Psi^{(0)}(z)+ \left[(r_0^-)^{2d-5}\omega\right]^{-\frac{d-3}{2d-5}}\Psi^{(1)}(z)\\
& \sim &
(A_+u_++A_-a_{--})\cos(z-\alpha_+)+(A_-u_-+A_+a_{++})\cos(z-\alpha_-)\\
&=&
\left[(A_+u_++A_-a_{--})e^{-i\alpha_+}+(A_-u_--A_+a_{++})e^{-i\alpha_-}\right]\frac{e^{iz}}2+\\&
&\left[(A_+u_++A_-a_{--})e^{i\alpha_+}+(A_-u_--A_+a_{++})e^{i\alpha_-}\right]\frac{e^{-iz}}2
\end{eqnarray*}
as $z \rightarrow \infty$.\\
 The function $\Psi_{\pm}^{(1)}$ defined
in (\ref{solutionrnds}) follows that
$$
\Psi_{\pm}^{(1)}=z^{1\pm j/2+\eta}G_{\pm}(z),
$$
where $\eta=-1/(4d-10)$, and $G_{\pm}(z)$ are even analytic
functions of $z$. By going around an arc of angle of
$\frac{2\pi}{2d-5}$ in complex $r$--plane (rotating from $A$ to the
next branch), $z$ rotates through an angle of $2\pi$ in $z$--plane,
leading to the wave function
\begin{eqnarray}
\Psi(z) &\sim &
\nonumber\left[(A_+v_--A_-a_{--}e^{-2ij\pi})e^{3i\alpha_+}+(A_-v_++A_+a_{++}e^{2ij\pi})e^{3i\alpha_-}\right]\frac{e^{2\eta\pi
i}e^{iz}}2+\\&&
\left[(A_+v_--A_-a_{--}e^{-2ij\pi})e^{5i\alpha_+}+(A_-v_++A_+a_{++}e^{2ij\pi})e^{5i\alpha_-}\right]\frac{e^{2\eta\pi
i}e^{-iz}}2.\label{rnds3}
\end{eqnarray}
as $z \rightarrow -\infty$. Here we have defined
$v_{\pm}=e^{-2\eta\pi i}\pm a_{+-}$.

As one follows the contour around the inner horizon $r=r_-$, the
wave function will be of the form
\begin{equation*}
\Psi(z-\delta) =
\Psi^{(0)}(z-\delta)+\left[(r_0^-)^{2d-5}\omega\right]^{-\frac{d-3}{2d-5}}\Psi^{(1)}(z-\delta),
\end{equation*}
where
$$
\delta = \frac{2\omega\pi i}{f'(r_-)} = \frac{\omega\pi i}{k_-},
$$
and $ k_- = \frac12 f'(r_-) $ is the surface gravity at the inner
horizon. Notice that $z-\delta< 0$ on this branch. As one lets
$z-\delta\rightarrow-\infty$, the wave function becomes
\begin{eqnarray}
\nonumber\Psi(z-\delta) &\sim &
\left[(B_+u_++B_-a_{--})e^{i\alpha_+}+(B_-u_--B_+a_{++})e^{i\alpha_-}\right]\frac{e^{i(z-\delta)}}2+\\
&&\left[(B_+u_++B_-a_{--})e^{-i\alpha_+}+(B_-u_--B_+a_{++})e^{-i\alpha_-}\right]\frac{e^{-i(z-\delta)}}2,\label{rnds4}
\end{eqnarray}
where we used
\begin{equation*}
J_\nu(z)=\sqrt{\frac{2}{\pi z}} \cos\left(z+\frac{\nu
\pi}{2}+\frac{\pi}4 \right) , \qquad z \ll -1.
\end{equation*}
Again, since on the same branch, we must let
$\Psi(z-\delta)=\Psi(e^{2\pi i}z)$. Then it is easily seen from
Eqs.(\ref{rnds3}) and (\ref{rnds4}) that we must have
\begin{eqnarray}
&& \nonumber\left[(A_+v_--A_-a_{--}e^{-2ij\pi})e^{3i\alpha_+}+(A_-v_++A_+a_{++}e^{2ij\pi})e^{3i\alpha_-}\right]e^{2\eta\pi i}=\\
&& \left[(B_+u_++B_-a_{--})e^{i\alpha_+}+(B_-u_--B_+a_{++})e^{i\alpha_-}\right]e^{-i\delta},\label{BB1}\\
&& \nonumber\left[(A_+v_--A_-a_{--}e^{-2ij\pi})e^{5i\alpha_+}+(A_-v_++A_+a_{++}e^{2ij\pi})e^{5i\alpha_-}\right]e^{2\eta\pi i}=\\
&&\left[(B_+u_++B_-a_{--})e^{-i\alpha_+}+(B_-u_--B_+a_{++})e^{-i\alpha_-}\right]e^{i\delta}.
\end{eqnarray}
Finally, we should rotate to the branch containing point $B$. This
make $z - \delta$ rotate through an angle of $2\pi$, leading to the
wave function
\begin{eqnarray*}
\Psi(z-\delta)&\sim &
\nonumber\left[(B_+v_--B_-a_{--}e^{-2ij\pi})e^{5i\alpha_+}+(B_-v_++B_+a_{++}e^{2ij\pi})e^{5i\alpha_-}\right]\frac{e^{2\eta\pi
i}e^{i(z-\delta)}}2+\\&&
\left[(B_+v_--B_-a_{--}e^{-2ij\pi})e^{3i\alpha_+}+(B_-v_++B_+a_{++}e^{2ij\pi})e^{3i\alpha_-}\right]\frac{e^{2\eta\pi
i}e^{-i(z-\delta)}}2.
\end{eqnarray*}
Consequently, we have
\begin{eqnarray}
\frac{(B_+v_--B_-a_{--}e^{-2ij\pi})e^{5i\alpha_+}+(B_-v_++B_+a_{++}e^{2ij\pi})e^{5i\alpha_-}}
{(A_+u_++A_-a_{--})e^{-i\alpha_+}+(A_-u_--A_+a_{++})e^{-i\alpha_-}}e^{2\eta\pi
i}e^{-i\delta} e^{\frac{\pi\omega}{k_+}+\frac{\pi\omega}{k_C}}& =& \mathcal{M}(r_+)\mathcal{M}(r_C),\label{7rnds}\\
\frac{(B_+v_--B_-a_{--}e^{-2ij\pi})e^{3i\alpha_+}+(B_-v_++B_+a_{++}e^{2ij\pi})e^{3i\alpha_-}}
{(A_+u_++A_-a_{--})e^{i\alpha_+}+(A_-u_--A_+a_{++})e^{i\alpha_-}}e^{2\eta\pi
i}e^{i\delta} e^{-\frac{\pi\omega}{k_H}-\frac{\pi\omega}{k_C}}& =&
\mathcal{M}(r_+)\mathcal{M}(r_C).\label{AB1}
\end{eqnarray}
The condition for these equations (\ref{BB1}$\sim$ \ref{AB1}) to
have nontrivial solutions $(A_+,A_-,B_+,B_-)$ is then
\begin{align*}
\left|
\begin{array}{ccccccc}
s_1e^{3i\alpha_+} & & s_7e^{3i\alpha_-} & & -s_6e^{i(\alpha_+-\delta)} & & -s_8e^{i(\alpha_--\delta)} \\
s_3e^{5i\alpha_+} & & s_5e^{5i\alpha_-} & & -s_2e^{-i(\alpha_+-\delta)} & & -s_4e^{-i(\alpha_--\delta)} \\
s_2e^{-i\alpha_+}e^{-\frac{2\pi \omega}{k_C}} & &
s_4e^{-i\alpha_-}e^{-\frac{2\pi \omega}{k_C}} & &
-s_3e^{i(5\alpha_+-\delta)} & & -s_5e^{i(5\alpha_--\delta)} \\
s_6e^{i\alpha_+}e^{\frac{2\pi \omega}{k^+}} & &
s_8e^{i\alpha_-}e^{\frac{2\pi \omega}{k^+}} & &
-s_1e^{i(3\alpha_++\delta)} & & -s_7e^{i(3\alpha_-+\delta)}
\end{array}
\right| = 0.
\end{align*}
where
\begin{eqnarray*}
s_1 &=& (v_-+a_{++}e^{\frac{ij\pi}2})e^{2\eta\pi i},\ \ \ s_2=u_+-a_{++}e^{\frac{ij\pi}2},\\
s_3 &=& (v_-+a_{++}e^{-\frac{ij\pi}2})e^{2\eta\pi i},\ \ \ s_4=u_-+a_{--}e^{-\frac{ij\pi}2},\\
s_5 &=& (v_+-a_{--}e^{\frac{ij\pi}2})e^{2\eta\pi i},\ \ \ s_6=u_+-a_{++}e^{-\frac{ij\pi}2},\\
s_7 &=& (v_+-a_{--}e^{-\frac{ij\pi}2})e^{2\eta\pi i},\ \ \
s_8=u_-+a_{--}e^{\frac{ij\pi}2}.
\end{eqnarray*}
After some algebra, we obtain
\begin{equation}\label{rnds1st}
\mathcal{A}_1\cosh\left(\frac{\pi\omega}{k_+}-\frac{\pi\omega}{k_C}-\Delta_1\right)
+\mathcal{A}_2\cosh\left(\frac{\pi\omega}{k_+}+\frac{2\pi\omega}{k_-}+\frac{\pi\omega}{k_C}+\Delta_2\right)+
\mathcal{A}_3\cosh(\frac{\pi\omega}{k_+}+\frac{\pi\omega}{k_C})=0,
\end{equation}
where
\begin{eqnarray*}
\mathcal{A}_1 &=&
\left[\left(2s_1s_3s_5s_7-s_3^2s_7^2e^{ij\pi}-s_1^2s_5^2e^{-ij\pi}\right)
\left(2s_2s_4s_6s_8-s_4^2s_6^2e^{ij\pi}-s_2^2s_8^2e^{-ij\pi}\right)\right]^{\frac12},\\
\mathcal{A}_2 &=&
\left[\left(2s_3s_5s_6s_8-s_3^2s_8^2e^{2ij\pi}-s_5^2s_6^2e^{-2ij\pi}\right)
\left(2s_1s_2s_4s_7-s_1^2s_4^2e^{2ij\pi}-s_2^2s_7^2e^{-2ij\pi}\right)\right]^{\frac12},\\
\mathcal{A}_3 &=& s_3s_4s_6s_7e^{ij\pi}+s_1s_2s_5s_8e^{-ij\pi}-
s_1s_3s_4s_8e^{2ij\pi}-s_2s_5s_6s_7e^{-2ij\pi},\\
\Delta_1 &=& \frac12
\left[\log\left(2s_1s_3s_5s_7-s_3^2s_7^2e^{ij\pi}-s_1^2s_5^2e^{-ij\pi}\right)-
\log\left(2s_2s_4s_6s_8-s_4^2s_6^2e^{ij\pi}-s_2^2s_8^2e^{-ij\pi}\right)\right],\\
\Delta_2 &=& \frac12
\left[\log\left(2s_3s_5s_6s_8-s_3^2s_8^2e^{2ij\pi}-s_5^2s_6^2e^{-2ij\pi}\right)-
\log\left(2s_1s_2s_4s_7-s_1^2s_4^2e^{2ij\pi}-s_2^2s_7^2e^{-2ij\pi}\right)\right].
\end{eqnarray*}
For zeroth-order asymptotic QN frequencies, one has $a_{\pm\pm}=0$,
and hence $u_{\pm}=1$ and $v_{\pm}=e^{-2\eta\pi i}$. As a result,
formula (\ref{rnds1st}) reduces to (\ref{rndsresult1}).

As discussed in Ref. \cite{natario-schiappa}, the Stokes lines for
$d=4$ and $d=5$ are a bit different from the case discussed in the
previous calculation for $d=6$. On account of the behavior of $r\sim
\infty$, they found that the final result for $d=4$ is the same as
Eq.(\ref{rnds1st}). However, the result for $d=5$ is changed. For
$r\sim\infty$, the coefficient of $e^{-iz}$ in the formula of the
wave function keeps unchanged, while the coefficient of $e^{iz}$ in
there reverses sign as one rotates from the branch containing point
$B$ to the branch containing point $A$ in the contour. Therefore,
one can complete this calculation by reversing the sign of Eq.
(\ref{7rnds}). In the end, we obtain
\begin{equation*}
\mathcal{A}_1\sinh\left(\frac{\pi\omega}{k_C}-\frac{\pi\omega}{k_+}+\Delta_1\right)
+\mathcal{A}_2\sinh\left(\frac{\pi\omega}{k_+}+\frac{2\pi\omega}{k_-}+\frac{\pi\omega}{k_C}+\Delta_2\right)+
\mathcal{A}_3\sinh(\frac{\pi\omega}{k_+}+\frac{\pi\omega}{k_C})=0,
\end{equation*}

For $d=4$ RNdS black holes, one has $r_C \gg r_+>r_-$ as $0<\lambda
\ll 1$. As a result, one has $-k_->k_+\gg k_C$. In this case, we can
neglect $1/k_+$ and $-1/k_-$ compared to $1/k_C$.  Taking Eq.
(\ref{rndsresult1}) into account, one finds the asymptotic QN
frequencies in this limit is the solutions of this formula
$$
e^{\frac{\pi\omega}{k_C}}+e^{-\frac{\pi\omega}{k_C}}=0.
$$
So, we have$ \frac{\omega}{T_C}=(2n+1)\pi i$, where $T_C$ is the
Hawking temperature at cosmological horizon. However, as one lets
$\lambda$ approaches its extremal value, \textit{i.e.},
$$\lambda\rightarrow
\frac{2+2\sqrt{9-8 q^2}}{(3+\sqrt{9-8 q^2})^3},
$$
with $ m=1$. \cite{molina} showed that in this extremal case we have
$k_H \approx k_C\rightarrow 0$, while $k_-$ does not approach zero.
So, in this limit, one can neglect the term with regard to $1/k_-$,
and obtain the asymptotic QN frequencies by solving this formula
$$
e^{2\frac{\pi\omega}{k_C}}+e^{-2\frac{\pi\omega}{k_C}}=0,
$$
which in turn derives $ \frac{\omega}{T_C}=\frac12(2n+1)\pi i$. This
reminds us to conjecture that the frequencies have values between
these two extremal values, \textit{i.e.},
\begin{equation}\label{chi2}
\frac{\omega}{T_C}\simeq \chi(2n+1)\pi i+{\mathbb{R}}{\mathrm{e}}\,
\omega,\ \ \ \ \ \frac12<\chi<1,
\end{equation}
where $\chi$ is a parameter closely related to the cosmological
constant $\lambda$. In fact, it would be an interesting work to
investigate the relationship of these two parameter, by analytical
methods or numerical ones. ${\mathbb{R}}{\mathrm{e}}\, \omega$ is
the real part of the frequency.

In order to obtain the explicit expressions of $\mathcal{A}_1$,
$\mathcal{A}_2$, $\Delta_1$ and $\Delta_2$, we need the integral
$$\int_0^{\infty}dz z^{-2/3}
J_{\nu}(z)J_{\mu}(z)=\frac{\Gamma(\frac23)\Gamma(\frac{
\mu+\nu+1/3}2)}{\sqrt[3]{4}\Gamma(\frac{\nu-\mu+5/3}2) \Gamma(\frac{
\mu+\nu+5/3}2)\Gamma(\frac{\mu-\nu+5/3}2)}.
$$
As a result, we obtain
\begin{eqnarray*}
s_1 &=& 1-2U(j)\sin\frac{(5+3j)\pi}6,\ \ \ \ \ \ \ \ \ \ \ \ \ \ \
s_2 = 1+(1-\sqrt{3}i)U(j)\sin\frac{(5+3j)\pi}6,\\
s_3 &=& 1+(1+\sqrt{3}i)U(j)\sin\frac{(5+3j)\pi}6,\ \ \ \
s_4 = 1+(1-\sqrt{3}i)U(j)\sin\frac{(5-3j)\pi}6,\\
s_5 &=& 1+(1+\sqrt{3}i)U(j)\sin\frac{(5-3j)\pi}6,\ \ \ \
s_6 = 1+2U(j)\sin\frac{(5+3j)\pi}6,\\
s_7 &=& 1-2U(j)\sin\frac{(5-3j)\pi}6,\ \ \ \ \ \ \ \ \ \ \ \ \ \ \
s_8 = 1+2U(j)\sin\frac{(5-3j)\pi}6,
\end{eqnarray*}
where
$$
U(j)=\frac{W(j)\Gamma(2/3)
\Gamma^2(1/6)\Gamma(1/6+j/2)\Gamma(1/6-j/2)}{16\pi^{2}r_0^-\sqrt{(4n+2)\chi
k_C}}.
$$
After some algebra, we obtain the explicit expressions of
$\mathcal{A}_1$, $\mathcal{A}_2$, $\Delta_1$ and $\Delta_2$ in terms
of $o(1/\sqrt{n})$
\begin{eqnarray*}
\mathcal{A}_1 &\simeq& 2(1-\cos j\pi)[1+(1+\sqrt{3}i)U(j)\cos j\pi/2],\\
\mathcal{A}_2 &\simeq& 2(1-\cos 2j\pi)\left\{1+\left[\cos j\pi/2-(3/2+\sqrt{3}i)\frac{\cos j\pi}{\cos j\pi/2}\right]U(j)\right\},\\
\mathcal{A}_3 &\simeq& 2(\cos j\pi-\cos 2j\pi)\left\{1+\left[4\cos j\pi/2+2\sqrt{3}i \frac{2\cos^2 j\pi-\cos j\pi}{1+2\cos j\pi}\right]U(j)\right\},\\
\Delta_1 &\simeq& (3\sqrt{3}i-5)U(j)\cos j\pi/2,\\
\Delta_2 &\simeq& U(j)\left[(2+\sqrt{3}i)\cos
j\pi/2+\frac{\sqrt{3}i}2\cdot\frac{\cos j\pi}{\cos j\pi/2}\right].
\end{eqnarray*}

For vector type perturbations, we have $j=5/3=2-1/3$. Therefore, as
one inserts $j=5/3$ into $\mathcal{A}_1$, $\mathcal{A}_2$,
$\mathcal{A}_3$, $\Delta_1$ and $\Delta_2$, we have exactly the same
results as in the $j=1/3$ case(tensor or scalar type perturbations),
and consequently Eq. (\ref{rnds1st}) becomes
\begin{eqnarray}\label{rnds1st4}
&&\nonumber\cosh\left(\frac{\pi\omega}{k_+}-\frac{\pi\omega}{k_C}+(5\sqrt{3}/2-3i)U(j)\right)
+\left[3-\frac{(3\sqrt{3}+15i)U(j)}2\right]\cosh\left(\frac{\pi\omega}{k_+}+\frac{2\pi\omega}{k_-}+\frac{\pi\omega}{k_C}+(\sqrt{3}+2i)U(j)\right)\\
&&+[2+(5\sqrt{3}+3i)U(j)]\cosh(\frac{\pi\omega}{k_+}+\frac{\pi\omega}{k_C})=0,
\end{eqnarray}
for three type perturbations. Inserting the expressions of $W(j)$
(see Appendix \ref{appendixA}), we can easily obtain the explicit
expression of $U(j)$. In this way, the first-order correction to
asymptotic QN frequencies of RNdS black holes can be obtained by
evaluating Eq. (\ref{rnds1st4}). The results show that the
correction term is closely related to $\ell$, $ q$, $\lambda$
(though the parameter $\chi$ and $k_C$), and of course, $n$.

Numerical calculation of this case has not been performed in the
previous literature. Consequently, it is interesting to perform
numerical checks to our first-order corrected results both in four
and higher dimensional black hole spacetime.

%%%%%%%%%%%%%%%%%%%%%%%%%%%%%%%%%%%%%%%%%%%%%%%%%%%%%%%%%%%%%%%%%

\subsection{The Schwarzschild Anti--de Sitter Case}

%%%%%%%%%%%%%%%%%%%%%%%%%%%%%%%%%%%%%%%%%%%%%%%%%%%%%%%%%%%%%%%%%

This case was studied analytically in \cite{musiri-siopsis1}. Here
we list their results for completeness. Considering the behavior of
the black hole singularity ($r \sim 0$) and large $r$, they obtained
the QN frequencies for all type of perturbations,
\begin{equation*}
\bar{z}=\frac{\pi}4(2+j+j_{\infty})-\tan^{-1}\frac
i{1+2e^{ij\pi}}+n\pi,
\end{equation*}
where $j_{\infty}=d-1, d-3$ and $d-5$ for tensor, vector and scalar
perturbations, respectively, and $\bar{z}=\omega\bar{r}_*$ is the
integration constant of the tortoise coordinate (refer the reader to
appendix \ref{appendixA}).

By expanding the wave function to the first order in
$1/\omega^{(d-3)/(d-2)}$ as
$$
\Psi=\Psi^{(0)}+\frac 1 {\omega^{(d-3)/(d-2)}} \Psi^{(1)},
$$
Ref. \cite{musiri-siopsis1} showed that the first-order correction
to the QN frequency is
\begin{equation*} \bar{z}=\omega
\bar{r}_*=\frac{\pi}4(2+j+j_{\infty})+\frac{\ln2}{2i}+n\pi-\frac18\left[6i\bar{b}-2ie^{-i\pi\frac{d-3}{d-2}}
\bar{b}-9\bar{a}_1+\bar{a}_2+e^{-i\pi\frac{d-3}{d-2}}(\bar{a}_1-\bar{a}_2)\right],
\end{equation*}
where
$$
\bar{a}_1=a_1(\infty),\ \ \ \bar{a}_2=a_2(\infty),\ \ \
\bar{b}=b(\infty),
$$
and $a_1(z), a_2(z), b(z)$ are defined as
\begin{eqnarray*}
 a_1(z) &=& \frac{\pi W(j)}8\omega^{-\frac{d-3}{d-2}}\int_0^zdz^{\prime}z^{\prime}{}^{-\frac1{d-2}}J_{j/2}(z^{\prime})
J_{j/2}(z^{\prime}),\\
a_2(z) &=& \frac{\pi
W(j)}8\omega^{-\frac{d-3}{d-2}}\int_0^zdz^{\prime}z^{\prime}{}^{-\frac1{d-2}}N_{j/2}(z^{\prime})
N_{j/2}(z^{\prime}),\\
b(z) &=& \frac{\pi
W(j)}8\omega^{-\frac{d-3}{d-2}}\int_0^zdz^{\prime}z^{\prime}{}^{-\frac1{d-2}}J_{j/2}(z^{\prime})
N_{j/2}(z^{\prime}).
\end{eqnarray*}
$W(j)$ in these formulae is defined as
\begin{eqnarray*}
W(j)=
 \begin{cases}
  W_{SAdST} &\text{$j=j_T$},\\
  W_{SAdSV} &\text{$j=j_{V}$},\\
  W_{SAdSS} &\text{$j=j_{S}$},
 \end{cases}
\end{eqnarray*}
 and the explicit expressions of $W_{SAdST}$, $W_{SAdSV}$ and $W_{SAdSS}$ can be found in appendix \ref{appendixA}.

For $d=4$, one obtains
\begin{equation}\label{corrsads}
\omega_n\bar{r}_*=\frac{\pi}4(2+j+j_{\infty})+n\pi+corr_4,
\end{equation}
where
$$
corr_4=\frac{\ln2}{2i}-\frac{(1+i)W(j)}
{128\pi^2}\cdot\sqrt{\frac{\bar{r}_*}{n}}\left[\cot\frac{\pi}2(\frac12-j)-3\right](\cos\frac{j\pi}2-\sin{j\pi}2)\Gamma^2(1/4)\Gamma(1/4+j/2)
\Gamma(1/4-j/2),
$$
and we have used the integral
$$
\int_0^{\infty}dz z^{-1/2} J_{\nu}(z)J_{
\mu}(z)=\frac{\Gamma(\frac12)\Gamma(\frac{\mu+\nu+1/2}2)}{\sqrt
{2}\Gamma(\frac{\nu-\mu+3/2}2)\Gamma(\frac{
\mu+\nu+3/2}2)\Gamma(\frac{\mu-\nu+3/2}2)}.
$$
In this way, one can easily obtain the explicit expression of
$corr_4$ of Eq. (\ref{corrsads}). \cite{musiri-siopsis1} have given
 the explicit expressions for tensor, vector and scalar type
perturbations in $d=4$ case. Moreover, they showed that their
results are in good agreement with the numerical results performed
in \cite{ckl}.

%%%%%%%%%%%%%%%%%%%%%%%%%%%%%%%%%%%%%%%%%%%%%%%%%%%%%%%%%%%%%%%%%

\subsection{The Reissner--Nordstr\"om Anti--de Sitter Case}

%%%%%%%%%%%%%%%%%%%%%%%%%%%%%%%%%%%%%%%%%%%%%%%%%%%%%%%%%%%%%%%%%
 We now compute analytically the quasinormal modes
of the RN AdS $d$--dimensional black hole including first-order
corrections. Previous work on this case can be found in
\cite{gubser-mitra-1,gubser-mitra-2,hubeny-rangamani}. Here we start
with the zeroth-order calculation. For Schwarzschild Anti--de Sitter
black hole, we have
\begin{equation}\label{fr}
f(r)=1-\frac{2 m}{r^{d-3}}-\lambda r^2,
\end{equation}
with the roots
$$
r_n=r_+, r_- r_1, r^*_1,\cdots, r_{d-3}, r^*_{d-3},
$$
where $\lambda (<0)$ is the black hole background parameter related
to the cosmological constant $\Lambda$ by
$$
\Lambda=\frac12 (d-1)(d-2)\lambda,
$$
and $r^*_n$ represents the conjugate of $r_n$. Near the black hole
singularity ($r \sim 0$), the tortoise coordinate may be expanded as
$$
r_*=\int \frac{dr}{f(r)}=\frac{1}{2d-5}\frac{r^{2d-5}}{ q^2}+\frac{2
m}{3d-8}\frac{r^{3d-8}}{ q^4}+\cdots.
$$
In this procedure, we have assumed $\frac r{r_0^-}\ll 1$, where
$r_0^-=\left( m-\sqrt{ m^2- q^2}\right)^{\frac1{d-3}}$, represents
the inner horizon of the RN black hole. Again we must expand the
potential to the first order in
$1/\left[(r_0^-)^{2d-5}\omega\right]^{(d-3)/(2d-5)}$ instead of
$1/\omega^{(d-3)/(2d-5)}$. After defining $z=\omega r_*$, the
potential for the three different type perturbations can be
expanded, respectively, as (appendix \ref{appendixA})
\begin{equation}\label{potential-rnads}
V [ z ] \sim -\frac{\omega^2}{4z^2} \left \{1-j^2-W(j)
\left(\frac{z}{(r_0^-)^{2d-5}\omega}\right)^{(d-3)/(d-2)}+\cdots
\right\} ,
\end{equation}
where
\begin{eqnarray*}
W(j)=
 \begin{cases}
  W_{RNAdST} &\text{$j=j_T$},\\
  W_{RNAdSV^{\pm}} &\text{$j=j_{V^{\pm}}$},\\
  W_{RNAdSS^{\pm}} &\text{$j=j_{S^{\pm}}$},
 \end{cases}
\end{eqnarray*}
 and the explicit expressions of $W_{RNAdST}$, $W_{RNAdSV^{\pm}}$ and $W_{RNAdSS^{\pm}}$ can be found in appendix \ref{appendixA}.
Then the Schr\"{o}dinger-like wave equation (\ref{schrodinger}) with
the potential (\ref{potential-rnads}) can be depicted as
\begin{equation}\label{waveequationrnads}
 \left(\mathcal{H}_0
+\left[(r_0^-)^{2d-5}\omega\right]^{-\frac{d-3}{d-2}}\mathcal{H}_1\right)
\Psi=0,
\end{equation}
 where
$\mathcal{H}_0$ and $\mathcal{H}_1$ are defined as
\begin{eqnarray*}
\mathcal{H}_0=\frac{d^2}{dz^2}+\left[\frac{1-j^2}{4z^2}+1\right] , \
\ \ \mathcal{H}_1=-\frac{W(j)}{4}z^{-\frac{3d-7}{2d-5}}.
\end{eqnarray*}
Obviously, the zeroth-order wave equation can be written as
$$
\mathcal{H}_0 \Psi^{(0)}=0,
$$
with solutions in the form of Bessel functions
$$
\Psi^{(0)}(z)=\sqrt{\frac{\pi z}2}J_{j/2}(z)+A\sqrt{\frac{\pi
z}2}N_{j/2}(z).
$$
As one lets $z\rightarrow +\infty$, it behaves as
\begin{equation*}
\Psi^{(0)}(z)\sim
\left(e^{-i\alpha_+}-iAe^{-i\alpha_+}\right)\frac{e^{iz}}2+
\left(e^{i\alpha_+}+iAe^{i\alpha_+}\right)\frac{e^{-iz}}2.
\end{equation*}
\FIGURE[ht]{\label{StokesRNAdS}
    \centering
    \psfrag{A}{$A$}
    \psfrag{B}{$B$}
    \psfrag{R+}{$r_+$}
    \psfrag{R-}{$r_-$}
    \psfrag{g1}{$r_1$}
    \psfrag{og1}{$r^*_1$}
    \psfrag{g2}{$r_2$}
    \psfrag{og2}{$r^*_2$}
    \psfrag{g3}{$r_3$}
    \psfrag{og3}{$r^*_3$}
    \psfrag{Re}{$\re$}
    \psfrag{Im}{$\im$}
    \psfrag{contour}{contour}
    \psfrag{Stokes line}{Stokes line}
    \epsfxsize=.6\textwidth
    \leavevmode
    \epsfbox{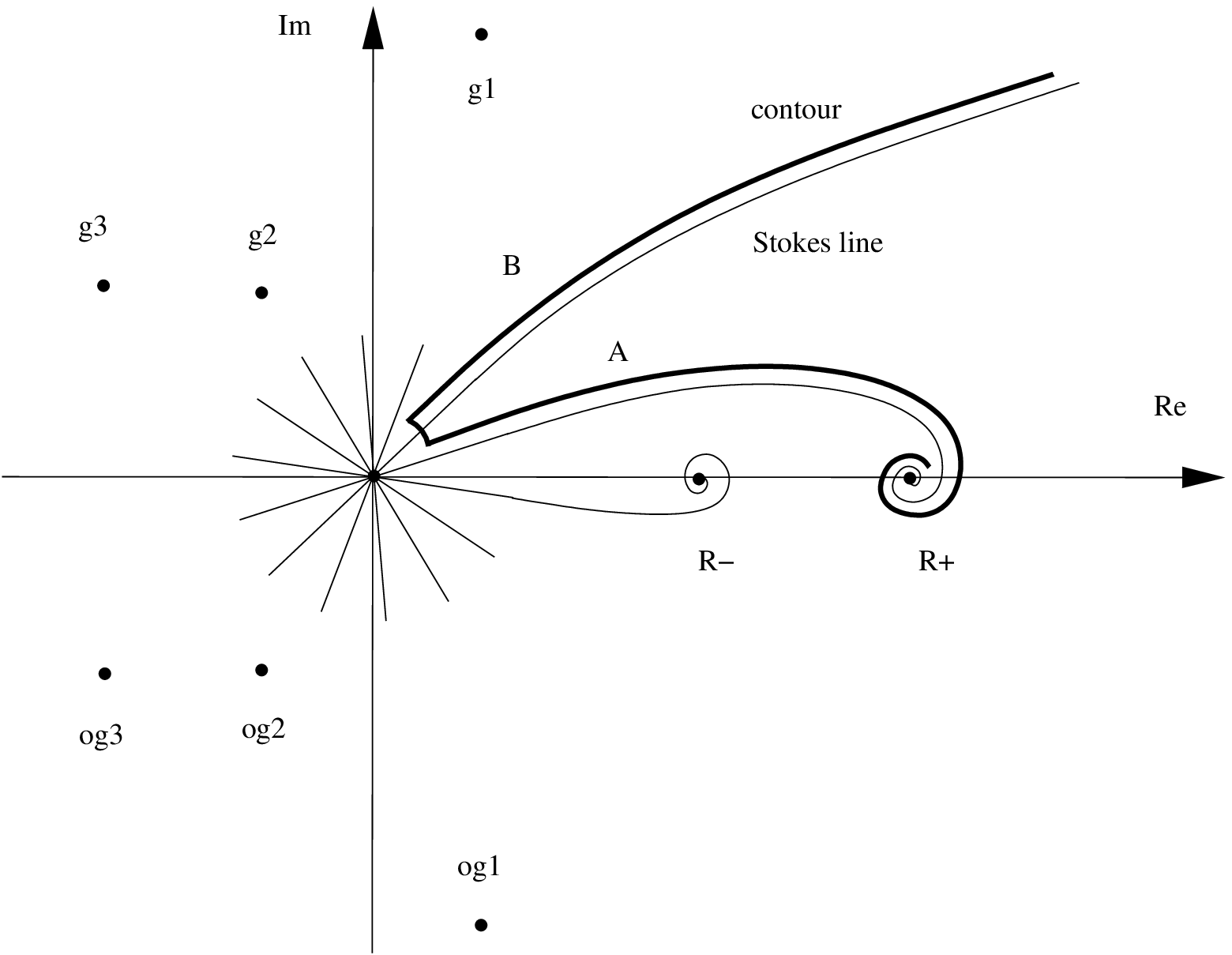}
\caption{Stokes line for the Reissner--Nordstr\"om Anti--de Sitter
black hole, along with the chosen contour for monodromy matching, in
the case of dimension $d=6$ ( we refer the reader to
\cite{natario-schiappa} for detail, and a more complete list of
figures in dimensions $d=4$, $d=5$, $d=6$ and $d=7$ ).} }
 This holds at point $B$ in Figure \ref{StokesRNAdS}. Appendix
\ref{appendixA} shows that the potential for large $r$ behaves as
\begin{equation*}
V=\frac{j_{\infty}-1}{4(z-\bar{z})^2}+\cdots
\end{equation*}
where $j_{\infty}=d-1,d-3$ and $d-5$ for tensor, vector and scalar
perturbations, respectively, and $\bar{z}$ is the integration
constant of the tortoise coordinate
$$
\bar{z}=\omega \bar{r}_*=\omega\int_0^{\infty}\frac{dr}{f(r)}.
$$
Then the Schr\"{o}dinger-like wave equation (\ref{schrodinger}) with
this potential can be depicted as
$$
\frac{d^2\Psi_{\infty}(z)}{dz^2}+\left[\frac{1-j_{\infty}^2}{4(z-\bar{z})^2}+1\right]\Psi_{\infty}(z)=0.
$$
On account of the boundary condition $\Psi_{\infty}(z)\rightarrow 0$
as $r\rightarrow\infty$, one can obtain the asymptotic behavior of
the wave function
\begin{equation}\label{rnadsinfin}
\Psi_{\infty}(z)\sim\frac{B}2\left[e^{iz}e^{-i(\bar{z}+\beta)}+e^{-iz}e^{i(\bar{z}-\beta)}\right],
\end{equation}
where $\beta= \frac{\pi}4(1+j_{\infty})$. Since on the same branch,
we must let $\Psi^{(0)}(z)=\Psi_{\infty}(z)$, then one finds
\begin{equation}\label{rnaads}
A=\tan(\bar{z}-\beta-\alpha_+).
\end{equation}
As one rotates from the branch containing point $B$ to the branch
containing point $A$ in the contour, through an angle of
$\frac{-\pi}{2d-5}$, $z \sim \frac{\omega r^{2d-5}}{(2d-5) q^2}$
rotates through an angle of $-\pi$, leading to the wave function
\begin{equation*}
\Psi^{(0)}(z) \sim \left(1-iA\right)\frac{e^{-i\alpha_+}e^{iz}}2+
\left[1-iA\left(1+2e^{ij\pi}\right)\right]\frac{e^{-3i\alpha_+}e^{-iz}}2,
\end{equation*}
as $z\rightarrow-\infty$.\\
From the boundary conditions (\ref{bounder}), we have
$$
1-iA\left(1+2e^{ij\pi}\right)=0.
$$
Consequently, taking Eq.(\ref{rnaads}) into account, one obtains the
QN frequencies
\begin{equation*}
\bar{z}=\frac{\pi}4(2+j+j_{\infty})-\tan^{-1}\frac
i{1+2e^{ij\pi}}+n\pi
\end{equation*}
Next, we calculate the first-order correction to the asymptotic
frequencies. Again we expand the wave function to the first order in
$1/\left[(r_0^-)^{2d-5}\omega\right]^{(d-3)/(2d-5)}$ as
$$
\Psi=\Psi^{(0)}+\frac 1
{\left[(r_0^-)^{2d-5}\omega\right]^{(d-3)/(2d-5)}} \Psi^{(1)}.
$$
Then one can rewrite Eq. (\ref{waveequationrnads}) as
\begin{equation*}
 \mathcal{H}_0
\Psi^{(1)}+\mathcal{H}_1 \Psi^{(0)}=0.
\end{equation*}
The general solution of this equation is
\begin{equation*}
\Psi^{(1)}(z)=\frac{\pi\sqrt{z}J_{j/2}(z)}2\int_0^zdz^{\prime}\sqrt{z^{\prime}}N_{j/2}(z^{\prime})
\mathcal{H}_1
\Psi^{(0)}(z^{\prime})-\frac{\pi\sqrt{z}N_{j/2}(z)}2\int_0^zdz^{\prime}\sqrt{z^{\prime}}J_{j/2}(z^{\prime})
\mathcal{H}_1 \Psi^{(0)}(z^{\prime}).
\end{equation*}
The behavior as $z\rightarrow +\infty$ is found to be
\begin{equation*}
\Psi_{\pm}^{(1)}(z) \sim \left(\mathcal{A}_1
e^{-i\alpha_+}-i\mathcal{A}_2 e^{-i\alpha_+}\right)\frac{e^{iz}}2+
\left(\mathcal{A}_1 e^{i\alpha_+}+i\mathcal{A}_2
e^{i\alpha_+}\right)\frac{e^{-iz}}2.
\end{equation*}
Here we defined
$$
\mathcal{A}_1=1-\bar{b}-A\bar{a}_2,\ \ \ \ \
\mathcal{A}_2=(1+\bar{b})A+\bar{a}_1,
$$
where
$$
\bar{a}_1=a_1(\infty),\ \ \ \bar{a}_2=a_2(\infty),\ \ \
\bar{b}=b(\infty),
$$
and $a_1(z), a_2(z), b(z)$ are defined as
\begin{eqnarray*}
 a_1(z) &=& \frac{\pi W(j)}8\left[(r_0^-)^{2d-5}\omega\right]^{-\frac{d-3}{2d-5}}\int_0^zdz^{\prime}z^{\prime}{}^{-\frac{d-2}{2d-5}}J_{j/2}(z^{\prime})
J_{j/2}(z^{\prime}),\\
a_2(z) &=& \frac{\pi
W(j)}8\left[(r_0^-)^{2d-5}\omega\right]^{-\frac{d-3}{2d-5}}\int_0^zdz^{\prime}z^{\prime}{}^{-\frac{d-2}{2d-5}}N_{j/2}(z^{\prime})
N_{j/2}(z^{\prime}),\\
b(z) &=& \frac{\pi
W(j)}8\left[(r_0^-)^{2d-5}\omega\right]^{-\frac{d-3}{2d-5}}\int_0^zdz^{\prime}z^{\prime}{}^{-\frac{d-2}{2d-5}}J_{j/2}(z^{\prime})
N_{j/2}(z^{\prime}).
\end{eqnarray*}
This holds at point $B$ in Figure \ref{StokesRNAdS}. For the same
reason have been discussed above for the zeroth-order case, we
should let $\Psi(z)=\Psi_{\infty}(z)$. Combining Eq.
(\ref{rnadsinfin}), one finds that
\begin{equation*}
\mathcal{A}_2=\mathcal{A}_1\tan(\bar{z}-\beta-\alpha_+)
\end{equation*}
as $z \rightarrow \infty$.

Again, as one rotates from the branch containing point $B$ to the
branch containing point $A$ in the contour, through an angle of
$\frac{-\pi}{2d-5}$, $z \sim \frac{\omega r^{2d-5}}{(2d-5) q^2}$
rotates through an angle of $-\pi$, leading to the wave function (as
$z\rightarrow-\infty$)
\begin{equation*}
\Psi^{(0)}(z) \sim \frac{Ce^{iz}}2+
\left\{\left[1-iA\left(1+2e^{ij\pi}\right)\right]e^{-3i\alpha_+}+e^{-i\alpha_+}e^{-i\pi\frac{d-3}{2d-5}}
\left(\bar{B}_1-i\bar{B}_2\right)\right\}\frac{e^{-iz}}2,
\end{equation*}
where
\begin{eqnarray*}
\bar{B}_1 &=&
-2\cos\frac{j\pi}2\left[e^{-ij\pi}\bar{a}_1+\left(A-e^{-2i\alpha_+}\right)\left[\bar{b}-i\left(1+e^{-ij\pi}\right)
\bar{a}_1\right]\right]-\\
&&-Ae^{-2i\alpha_+}\left[e^{ij\pi}\bar{a}_2-4\cos^2\frac{j\pi}2\bar{a}_1-2i\left(1+e^{ij\pi}\right)
\bar{b}\right],\\
\bar{B}_2 &=&
e^{-2i\alpha_-}\left[e^{-ij\pi}\bar{a}_1+A\left[\bar{b}-i\left(1+e^{-ij\pi}\right)\bar{a}_1\right]\right],
\end{eqnarray*}
and $C$ is a constant that can be easily calculated but is not needed here. \\
Taking the boundary conditions (\ref{bounder}) into account, we have
$$
\left[1-iA\left(1+2e^{ij\pi}\right)\right]e^{-3i\alpha_+}+e^{-i\alpha_+}e^{-i\pi\frac{d-3}{2d-5}}
\left(\bar{B}_1-i\bar{B}_2\right)=0.
$$
Consequently, combining Eq.(\ref{rnaads}), one obtains the QN
frequencies
\begin{equation*}
\bar{z}=\omega
\bar{r}_*=\frac{\pi}4(2+j+j_{\infty})+\frac{\ln2}{2i}+n\pi-\frac18\left[6i\bar{b}-2ie^{-i\pi\frac{d-3}{2d-5}}
\bar{b}-9\bar{a}_1+\bar{a}_2+e^{-i\pi\frac{d-3}{2d-5}}(\bar{a}_1-\bar{a}_2)\right]
\end{equation*}
For $d=4$, one obtains
\begin{equation}\label{corrrnads}
\omega_n\bar{r}_*=\frac{\pi}4(2+j+j_{\infty})-\tan^{-1}\frac
i{1+2e^{ij\pi}}+n\pi+corr_4,
\end{equation}
where
\begin{eqnarray*}
corr_4 &=& \frac{1}{2i}\ln(\cos
j\pi/2)-\frac{(1+\sqrt{3}i)W(j)}{1024
\pi^2r_0^-}\sqrt[3]{\frac{2\bar{r}_*}{n\pi}}
\left[(5\sqrt{3}+1)\sin j\pi/2-(5-\sqrt{3})\cos j\pi/2\right]\cdot\\
&&\cdot\Gamma(2/3)\Gamma^2(1/6)\Gamma(1/6+j/2) \Gamma(1/6-j/2),
\end{eqnarray*}
and we have used the integral
$$
\int_0^{\infty}dz z^{-2/3} J_{\nu}(z)J_{
\mu}(z)=\frac{\Gamma(\frac23)\Gamma(\frac{\mu+\nu+1/3}2)}{\sqrt[3]
{4}\Gamma(\frac{\nu-\mu+5/3}2)\Gamma(\frac{\mu+\nu+5/3}2)\Gamma(\frac{
\mu-\nu+5/3}2)}.
$$
For $d=4$, the roots of the formula $f(r)=0$ can be depicted by
$r_+$ (the black hole outer horizon), $r_-$ (the black hole inner
horizon), and two complex roots, $r_1$ and its complex conjugate
$r^*_1$. As discussed in \cite{musiri-siopsis1}, it is convenient to
set $\lambda=-1$. From (\ref{fr}), the roots are
\begin{eqnarray*}
r_+ &=& \frac12(g+\sqrt{-2-g^2+4 m/g}),\ \ \ \
r_- = \frac12(g-\sqrt{-2-g^2+4 m/g}),\\
r_1 &=& \frac12(-g+\sqrt{-2-g^2-4 m/g}),\ \ \ \  r^*_1 =
\frac12(-g-\sqrt{-2-g^2-4 m/g}),
\end{eqnarray*}
with $g=\sqrt{-\frac23+\frac{h_++h_-}{3\sqrt{2}}}$, where
$$
h_{\pm}=\left(2-72 q^2+108 m^2\pm\sqrt{-4(1+12 q^2)^3+(2-72 q^2+108
m^2)^2}\right)^{\frac13}.
$$
The integration constant in the tortoise coordinate can be then
solved as
\begin{eqnarray*}
\bar{r}_* &=& \int_0^{\infty}\frac{dr}{f(r)} \\
&=&-\frac{r_-^2}{3r_-^3+r_--r_+r_1r^*_1}\ln\frac{r_-}{r_+}
-\frac{r_1^2}{3r_1^3+r_1-r_+r_-r^*_1}\ln\frac{r_1}{r_+}
-\frac{r^*_1{}^2}{3r^*_1{}^3+r^*_1-r_+r_-r_1}\ln\frac{r^*_1}{r_+}.
\end{eqnarray*}
Obviously, $\bar{r}_*$ is the formula (105) in
\cite{musiri-siopsis1} as $ q=0$ (or the Schwarzschild limit).

In this way, one can easily obtain the value of QN frequencies in
Eq. (\ref{corrrnads}). For tensor type perturbations,
$W_{RNAdST}=0$, so $corr_4=-i\frac{\ln 3}{4}$. However, it seems
unavailable for scalar type perturbations, since in this case we
have $j\rightarrow1/3$, which may induce the integral
$\int_0^{\infty}dz z^{-2/3} J_{-1/3}(z)J_{-1/3}(z)$ approaches
infinity. It is interesting to investigate this problem. Is there
any other methods can avoid this singularity? For vector type
perturbations , we have $j\rightarrow5/3$. This leads to
$$
corr_4= -i\frac{\ln
3}{4}+\frac{(15-\sqrt{3})(1+\sqrt{3}i)W_{RNAdSV^{\pm}}}{1024 \pi
r_0^-}\sqrt[3]{\frac{2\bar{r}_*}{n\pi}} \cdot\Gamma^2(1/6),
$$
where
$$
W_{RNAdSV^{\pm}}=\frac43\left( m\pm\sqrt{9 m^2+4\ell(\ell+1) q^2-8
q^2}\right)(1-\sqrt{1- q^2})(3 q^2)^{-2/3}
$$
From this formula we find: (1) in order to insure
$r_0^-(4n+2)^{1/3}\gg 1$ as one calculates the QN frequencies of the
first-order correction, the imaginary part of the frequencies ( or
the modes $n$) needs bigger values for a black hole with small
charge, since $r_0^- \sim  q^2\rightarrow 0$ as $ q\rightarrow0$.
This confirms the prediction made by Neitzke in \cite{neitzke}: the
required $n$ diverges as $ q\rightarrow 0$, and the corrections
would blow up this divergence; (2) the first-order correction to the
asymptotic QN frequencies are shown to be dimension dependent and
related closely to $ m$, $\ell$, $j$, and the charge $ q$.

QN frequencies of RN Ads black holes were first calculated
numerically by Berti and Kokkotas in \cite{berti-kokkotas-1}, in the
case of $d=4$. \cite{wlm} latter performed an extensive numerical
study of QN frequencies for massless scalar fields in $d=4$ RN AdS
spacetime. They find for the higher modes, both
${\mathbb{R}}{\mathrm{e}}\, \omega$ and ${\mathbb{I}}{\mathrm{m}}\,
\omega$ increase with $n$. As $ q$ increases,
${\mathbb{I}}{\mathrm{m}}\, \omega$ increases faster than
${\mathbb{R}}{\mathrm{e}}\, \omega$ does, with $n$.
\cite{natario-schiappa} matched these predictions to their
zeroth-order analytical formula of the asymptotic QN frequency.
Their results are complete agreement with the numerical calculations
of \cite{wlm}. In our first-order corrected formula, the frequencies
can be obtained by subtracting $i\frac{\ln 3}4$ from their results.
This leads, obviously, to the same gap of $\omega$ as
\cite{natario-schiappa} does. Although more complete numerical
results are needed to check our analytical results in higher
dimensional cases, our results are in good agreement with the
numerical results in the particular case of $d=4$.

%%%%%%%%%%%%%%%%%%%%%%%%%%%%%%%%%%%%%%%%%%%%%%%%%%%%%%%%%%%%%%%%%
%%%%%%%%%%%%%%%%%%%%%%%%%%%%%%%%%%%%%%%%%%%%%%%%%%%%%%%%%%%%%%%%%

\section{Conclusions and Future Directions}

%%%%%%%%%%%%%%%%%%%%%%%%%%%%%%%%%%%%%%%%%%%%%%%%%%%%%%%%%%%%%%%%%
%%%%%%%%%%%%%%%%%%%%%%%%%%%%%%%%%%%%%%%%%%%%%%%%%%%%%%%%%%%%%%%%%

In this paper we studied analytically quasinormal modes in a wide
variety of black hole spacetimes, including $d$--dimensional
asymptotically flat spacetimes and non-asymptotically flat
spacetimes. We extend the procedure of \cite{natario-schiappa} to
include first-order corrections to analytical expressions for QN
frequencies by making use of a monodromy technique, which was first
introduced in \cite{motl-neitzke} for zeroth-order approximation,
and later extended to first-order for Schwarzschild black hole in
\cite{musiri-siopsis}. The calculation performed in this paper show
that systematic expansions for uncharged black holes include
different corrections with the one for charged black holes. In other
words, $d$--dimensional uncharged black holes have an expansion
including corrections in $1/\omega^{(d-2)/(d-3)}$, while charged
ones have an expansion in $1/\omega^{(d-3)/(2d-5)}$. This difference
makes them have a different $n$--dependence relation in the
first-order correction formulae.

The method applied above in calculating the first-order corrections
of QN frequencies seems to be unavailable for black holes with small
charge, since the required $n$ diverges as the charge $
q\rightarrow0$, and worse, the corrections would blow this
divergence up. More extensive investigation is needed for this
problem. A very recent work on RN black holes appeared in
\cite{dkob}, where they discussed a possible way to avoid this
divergence for zeroth-order case in the small charge limit.

Some remarks are due for the particular case of $d=4$ dimensional
spacetimes, as in these cases we discussed more deeply. For
Schwarzschild case, we obtained a $j$ and $\ell$--dependence
correction term, which was proved by numerical results. For RN case,
a puzzle appears when we apply our method to calculate the
first-order corrections of the scalar type perturbations, since the
correction in this case approaches infinity. In fact, this problem
also exists in other four dimensional charged black holes. Strangely
enough, extremal RN black hole can avoid this problem. Moreover, it
seems that the correction term equals to zero, independently of the
type perturbations. An investigation on whether the first-order
corrections for any $d>3$ extremal RN black holes have this
behavior, should be further performed. Another point worthy of being
mentioned about RN case is that the extremal limits ($
q\rightarrow0$, or $ q\rightarrow m$) of the RN QN frequencies do
not yield the right QN frequencies in their extremal cases
(Schwarzschild QN frequencies, or extremal RN QN frequencies). Our
results show that the same thing happens in the first-order
corrections. \cite{andersson-howls} found that the limit is a
singular one since it involves topology change at the level of the
contours in the complex plane.

On the applications to quantum gravity \cite{natario-schiappa}
 indicate that the $\ln3$ in $d=4$ Schwarzschild seems to be nothing
but some numerical coincidences. Our results support this argument.
However, a recent paper on quantization of charged black holes
\cite{hod2} found that other than the Schwarzschild case, the
asymptotic resonance corresponds to a fundamental area unit $\Delta
A=4\hbar \ln2$. According to Dreyer's conjecture \cite{dreyer}, one
can immediately set the spin network unit as $1/2$, \textit{i.e.},
$j_{min}=1/2$. This differs from $j_{min}=1$ for Schwarzschild case.
From the LQG point of view, this result support the claims
\cite{corichi} that the gauge group of LQG should be SU(2) in spite
of the $\ln3$ for Schwarzschild. Is there any other explanation of
that? Or more generally, can we find other similar proof to support
Dreyer's conjecture for other spacetimes?

There are some other possible directions for further study: (1)
perform some numerical calculations to check our analytical
solutions done above; (2) make sure if it is possible to have such
an easy relation between the asymptotic QN frequencies and the
overtone modes $n$ as suggested in the asymptotically dS
cases(\ref{chi1}) and (\ref{chi2}), or more deeply, find the
possible relation between $\chi$ and $\lambda$ and a possible
explanation of that; (3) find an explanation of our correction
results on the physical side.

\section*{Acknowledgements}

We would like to thank Emanuele Berti, Suphot Musiri, Andrew Neitzke
and George Siopsis, for their comments and discussions, at different
stages of this work, and specially Jos\'e Nat\'ario and Ricardo
Schiappa for their figures in this paper. The work was supported by
the National Natural Science Foundation of China under Grant No.
10573027, and the Foundation of Shanghai Natural Science Foundation
under Grant No. 05ZR14138.

\vfill

\eject

\appendix

%%%%%%%%%%%%%%%%%%%%%%%%%%%%%%%%%%%%%%%%%%%%%%%%%%%%%%%%%%%%%%%%%
%%%%%%%%%%%%%%%%%%%%%%%%%%%%%%%%%%%%%%%%%%%%%%%%%%%%%%%%%%%%%%%%%

\section{The Expanded Master Equation Potentials} \label{appendixA}

%%%%%%%%%%%%%%%%%%%%%%%%%%%%%%%%%%%%%%%%%%%%%%%%%%%%%%%%%%%%%%%%%
%%%%%%%%%%%%%%%%%%%%%%%%%%%%%%%%%%%%%%%%%%%%%%%%%%%%%%%%%%%%%%%%%

In this appendix we shall deduce the expressions of the potentials
in Schr\"odinger--like equation (\ref{schrodinger}). Nat\'{a}rio and
Schiappa \cite{natario-schiappa} have deduced the value of the
potentials in the region of large $r$. For completeness, we may list
them in this appendix. Our work focuses on the region of the origin,
which plays an key role in our procedure to calculate the
first-order correction of the asymptotic QN frequencies using
monodromy method.

We start with a set of potentials determined by the type of
perturbations and can be deduced from a set of master equations
which were derived by Ishibashi and Kodama in
\cite{kodama-ishibashi-1, kodama-ishibashi-2, kodama-ishibashi-3}
(can be denoted as the Ishibashi--Kodama master equations). Making
use of these potentials, we make an exhaustive study of the
first-order values of all these potentials in terms of the tortoise
coordinate. The zeroth-order values of these potentials can be found
in \cite{natario-schiappa}. For tensor, vector and scalar
perturbations, we obtain, respectively,

\begin{equation} \label{tensor}
V_{\mathsf{T}} (r) = f(r) \left( \frac{\ell \left( \ell+d-3
\right)}{r^{2}} + \frac{\left( d-2 \right) \left( d-4 \right)
f(r)}{4 r^{2}} + \frac{\left( d-2 \right) f'(r)}{2r} \right).
\end{equation}

\begin{equation} \label{vector-charged}
V_{\mathsf{V}^{\pm}} (r) =  f(r) \left( \frac{\ell \left( \ell+d-3
\right)}{r^{2}} + \frac{\left( d-2 \right) \left( d-4 \right)
f(r)}{4 r^{2}}  - \frac{\left( d-1 \right)  m}{r^{d-1}} +
\frac{\left( d-2 \right)^{2}  q^{2}}{r^{2d-4}} \pm
\frac{\Delta}{r^{d-1}} \right),
\end{equation}
where
\begin{equation}\label{delta}
\Delta \equiv \sqrt{\left( d-1 \right)^{2} \left( d-3 \right)^{2}
 m^{2} + 2 \left( d-2 \right) \left( d-3 \right) \Big( \ell \left(
\ell+d-3 \right) - \left( d-2 \right) \Big)  q^{2}}.
\end{equation}

\begin{equation} \label{scalar-charged}
V_{\mathsf{S}^{\pm}} (r) = \frac{f(r) U_{\pm} (r)}{64 r^{2}
H_{\pm}^{2} (r)}.
\end{equation}
The expressions of $U_{\pm}$ and $H_{\pm}$ can be found in
\cite{natario-schiappa}.
In these expressions, the $\Psi_{+}$
equation represents the electromagnetic mode and the $\Psi_{-}$
equation represents the gravitational mode.

\medskip

\noindent \underline{\textsf{The Schwarzschild Case:}} In this case,
the black hole is uncharged, so one has
$V_{\mathsf{V^+}}=V_{\mathsf{V^-}}$ and
$V_{\mathsf{S^+}}=V_{\mathsf{S^-}}$. For convenience, we denote them
by $V_{\mathsf{V}}$ and $V_{\mathsf{S}}$, respectively. The
background space-time metric is
$$
f(r)=1-\frac{2 m}{r^{d-3}}.
$$
Near the black hole singularity ($r \sim 0$), the tortoise
coordinate can be expanded as
$$
r_*=\int \frac{dr}{f(r)}=-\frac{1}{d-2}\frac{r^{d-2}}{2
m}-\frac{1}{2d-5}\frac{r^{2d-5}}{(2 m)^2}+\cdots,
$$
where we have kept the second term in the expansion of $r$ and have
chosen the integration constant so that $r_*=0$ at $r=0$. Using this
formula, one may obtain an approximate expression of $r$ in terms of
the tortoise coordinate $r_*$
\begin{equation}\label{apprors}
r\sim[-2 m (d-2)r_*]^{\frac1{d-2}}+\frac{(d-2)r_*}{2d-5}.
\end{equation}
As one applies (\ref{apprors}) to the three different potentials
(eqs. (\ref{tensor}), (\ref{vector-charged}) and
(\ref{scalar-charged})), these potentials can be expanded near the
black hole singularity in terms of the tortoise coordinate,
respectively, as
\begin{eqnarray*}
V_{\mathsf{T}} (r) &\sim& - \frac{1}{4 z^{2}}\left[1-0^2-W_{ST}\left(\frac{z}{\omega}\right)^{\frac{d-3}{d-2}}\right]\\
V_{\mathsf{V}} (r) &\sim& - \frac{1}{4 z^{2}}\left[1-2^2-W_{SV}\left(\frac{z}{\omega}\right)^{\frac{d-3}{d-2}}\right]  \\
V_{\mathsf{S}} (r) &\sim& - \frac{1}{4
z^{2}}\left[1-0^2-W_{SS}\left(\frac{z}{\omega}\right)^{\frac{d-3}{d-2}}\right],
\end{eqnarray*}
where
\begin{eqnarray*}
W_{ST} &=& \frac1{[-2(d-2)
m]^{\frac1{d-2}}}\left[\frac{2(d-3)^2}{2d-5}+\frac{4\ell(\ell+d-3)}{d-2}\right],
\\
W_{SV} &=& \frac1{[-2(d-2)
m]^{\frac1{d-2}}}\left[\frac{2(d^2-8d+13)}{2d-15}+\frac{4\ell(\ell+d-3)}{d-2}\right],
\\
W_{SS} &=& \frac1{[-2(d-2)
m]^{\frac1{d-2}}}\left[\frac{2d^3-24d^2+94d-116}{(2d-5)(d-2)}+\frac{4(d^2-7d+14)[\ell(\ell+d-3)-d+2]}{(d-1)(d-2)^2}\right],
\end{eqnarray*}
and we have rescaled the tortoise coordinate $z = \omega r_*$.

Near $r = \infty$ one has
\begin{eqnarray*}
V_{\mathsf{T}} (r) &\sim& 0, \\
V_{\mathsf{V}} (r) &\sim& 0, \\
V_{\mathsf{S}} (r) &\sim& 0.
\end{eqnarray*}

\medskip

\noindent \underline{\textsf{The RN Case:}} The background
space-time metric is
$$
f(r)=1-\frac{2 m}{r^{d-3}}+\frac{ q^2}{r^{2d-6}}.
$$
Near the black hole singularity ($r \sim 0$), the tortoise
coordinate can be expanded as
$$
r_*=\int \frac{dr}{f(r)}=\frac{1}{2d-5}\frac{r^{2d-5}}{ q^2}+\frac{2
m}{3d-8}\frac{r^{3d-8}}{ q^4}+\cdots.
$$
where we have kept the second term in the expansion of $r$ and have
chosen the integration constant so that $r_*=0$ at $r=0$, and we
have assumed $r/r_0^-\ll1$. Using this formula, one may obtain an
approximate expression of $r$ in terms of the tortoise coordinate
$r_*$
\begin{equation}\label{approrrn}
\frac r{r_0^-}\sim [(2d-5)
q^2(r_0^-)^{5-2d}r_*]^{\frac1{2d-5}}-\frac{2 m(r_0^-)^{d-3}}{(3d-8)
q^2}[(2d-5) q^2(r_0^-)^{5-2d}r_*]^{\frac{d-2}{2d-5}}.
\end{equation}
As one applies (\ref{approrrn}) to the five different potentials
(eqs. (\ref{tensor}), (\ref{vector-charged}) and
(\ref{scalar-charged})), these potentials can be expanded near the
black hole singularity in terms of the tortoise coordinate,
respectively, as
\begin{eqnarray*}
V_{\mathsf{T}} (r) &\sim& - \frac{1}{4 z^{2}}\left[1-j_T^2-W_{RNT}\left(\frac{z}{(r_0^-)^{2d-5}\omega}\right)^{\frac{d-3}{2d-5}}\right]\\
V_{\mathsf{V}^{\pm}} (r) &\sim& - \frac{1}{4 z^{2}}\left[1-j_{V^{\pm}}^2-W_{RNV^{\pm}}\left(\frac{z}{(r_0^-)^{2d-5}\omega}\right)^{\frac{d-3}{2d-5}}\right]  \\
V_{\mathsf{S}^{\pm}} (r) &\sim& - \frac{1}{4
z^{2}}\left[1-j_{S^{\pm}}^2-W_{RNS^{\pm}}\left(\frac{z}{(r_0^-)^{2d-5}\omega}\right)^{\frac{d-3}{2d-5}}\right],
\end{eqnarray*}
where
$$
j_T= j_{S^{\pm}}= \frac{d-3}{2d-5},\ \ \ \
j_{V^{\pm}}=\frac{3d-7}{2d-5}=2-j_T,
$$
and we have rescaled the tortoise coordinate $z = \omega r_*$.
The expressions of $W_{RNT}$, $W_{RNV^{\pm}}$ and $W_{RNS^{\pm}}$ in
these equations are
\begin{eqnarray*}
W_{RNT} &=& 0,
\\
W_{RNV^{\pm}} &=& \frac{4 m(d-2)(d^2-d-4)-4(3d-8)[(d-1)
m\mp\Delta]}{(2d-5)(3d-8)}(r_0^-)^{d-3}\left[(2d-5)
q^2\right]^{\frac{2-d}{2d-5}},
\\
W_{RNS^{\pm}} &=& -[2 m(d-2)(4d-10)
q^2+\mathcal{W}_{\pm}](r_0^-)^{d-3}\left[(2d-5)
q^2\right]^{\frac{7-3d}{2d-5}},
\end{eqnarray*}
where
\begin{eqnarray*}
\mathcal{W}_+ &=&
- m q^2[-(d-1)(3d-8)(1-\Omega)+2(4d^2-15d+12)]-2(d-2)(3d-8) m q^2-\\
&&-\frac{4(3d-8)[\ell(\ell+d-3)-(d-2)] q^4}{(d-1)(1-\Omega) m},\\
\mathcal{W}_- &=&
- m q^2[(d-1)(3d-8)(1-\Omega)+2(d-2)^2]-2(d-2)(3d-8) m q^2-\\
&&-\frac{4(3d-8)[\ell(\ell+d-3)-(d-2)] q^4}{(d-1)(1+\Omega) m},
\end{eqnarray*}
with the definition, here and below, that
\begin{equation*}
\Omega=\sqrt{1+\frac{4[\ell(\ell+d-3)-d+2]q^2}{(d-1)^2m^2}}.
\end{equation*}
For the extremal case, $q \rightarrow m$, we have
\begin{eqnarray*}
V_{\mathsf{T}} (r) &\sim& - \frac{1}{4 z^{2}}\left[1-j_T^2-W_{RNT}^{ex}\left(\frac{z}{\omega}\right)^{\frac{d-3}{2d-5}}\right]\\
V_{\mathsf{V}^{\pm}} (r) &\sim& - \frac{1}{4 z^{2}}\left[1-j_{V^{\pm}}^2-W_{RNV^{\pm}}^{ex}\left(\frac{z}{\omega}\right)^{\frac{d-3}{2d-5}}\right]  \\
V_{\mathsf{S}^{\pm}} (r) &\sim& - \frac{1}{4
z^{2}}\left[1-j_{S^{\pm}}^2-W_{RNS^{\pm}}^{ex}\left(\frac{z}{\omega}\right)^{\frac{d-3}{2d-5}}\right],
\end{eqnarray*}
where
\begin{eqnarray*}
W_{RNT}^{ex} &=& 0,
\\
W_{RNV^{\pm}}^{ex} &=& \frac{4 m(d-2)(d^2-d-4)-4(3d-8)[(d-1)
m\mp\Delta^{ex}]}{(2d-5)(3d-8)}\left[(2d-5)
m^2\right]^{\frac{2-d}{2d-5}},
\\
W_{RNS^{\pm}}^{ex} &=& 0,
\end{eqnarray*}
and
\begin{equation*}
 \Delta^{ex}=\sqrt{\left( d-1 \right)^{2}
\left( d-3 \right)^{2}
 + 2 \left( d-2 \right) \left( d-3 \right) \Big( \ell \left(
\ell+d-3 \right) - \left( d-2 \right) \Big) } m.
\end{equation*}

 Near $r = \infty$ one finds
\begin{eqnarray*}
V_{\mathsf{T}} (r) &\sim& 0, \\
V_{\mathsf{V}} (r) &\sim& 0, \\
V_{\mathsf{S}} (r) &\sim& 0.
\end{eqnarray*}

\medskip

\noindent \underline{\textsf{The Schwarzschild dS Case:}} In this
case, the black hole is uncharged, so one has
$V_{\mathsf{V^+}}=V_{\mathsf{V^-}}$ and
$V_{\mathsf{S^+}}=V_{\mathsf{S^-}}$. For convenience, we denote them
by $V_{\mathsf{V}}$ and $V_{\mathsf{S}}$, respectively. The
background space-time metric is
$$
f(r)=1-\frac{2 m}{r^{d-3}}-\lambda r^2,
$$
with $\lambda>0$. Near the black hole singularity ($r \sim 0$), the
tortoise coordinate can be expanded as
$$
r_*=\int \frac{dr}{f(r)}=-\frac{1}{d-2}\frac{r^{d-2}}{2
m}-\frac{1}{2d-5}\frac{r^{2d-5}}{(2 m)^2}+\cdots,
$$
where we have kept the second term in the expansion of $r$ and have
chosen the integration constant so that $r_*=0$ at $r=0$. Using this
formula, one may obtain an approximate expression of $r$ in terms of
the tortoise coordinate $r_*$
\begin{equation}\label{approrsds}
r\sim[-2 m (d-2)r_*]^{\frac1{d-2}}+\frac{(d-2)r_*}{2d-5}.
\end{equation}
As one applies (\ref{approrsds}) to the three different potentials
(eqs. (\ref{tensor}), (\ref{vector-charged}) and
(\ref{scalar-charged})), these potentials can be expanded near the
black hole singularity in terms of the tortoise coordinate,
respectively, as
\begin{eqnarray*}
V_{\mathsf{T}} (r) &\sim& - \frac{1}{4 z^{2}}\left[1-0^2-W_{SdST}\left(\frac{z}{\omega}\right)^{\frac{d-3}{d-2}}\right]\\
V_{\mathsf{V}} (r) &\sim& - \frac{1}{4 z^{2}}\left[1-2^2-W_{SdSV}\left(\frac{z}{\omega}\right)^{\frac{d-3}{d-2}}\right]  \\
V_{\mathsf{S}} (r) &\sim& - \frac{1}{4
z^{2}}\left[1-0^2-W_{SdSS}\left(\frac{z}{\omega}\right)^{\frac{d-3}{d-2}}\right],
\end{eqnarray*}
where
\begin{eqnarray*}
W_{SdST} &=& \frac1{[-2(d-2)
m]^{\frac1{d-2}}}\left[\frac{2(d-3)^2}{2d-5}+\frac{4\ell(\ell+d-3)}{d-2}\right],
\\
W_{SdSV} &=& \frac1{[-2(d-2)
m]^{\frac1{d-2}}}\left[\frac{2(d^2-8d+13)}{2d-15}+\frac{4\ell(\ell+d-3)}{d-2}\right],
\\
W_{SdSS} &=& \frac1{[-2(d-2)
m]^{\frac1{d-2}}}\left[\frac{2d^3-24d^2+94d-116}{(2d-5)(d-2)}+\frac{4(d^2-7d+14)[\ell(\ell+d-3)-d+2]}{(d-1)(d-2)^2}\right],
\end{eqnarray*}
and we have rescaled the tortoise coordinate $z = \omega r_*$. It is
easily seen that this is just like in the pure Schwarzschild case.

Near $r = \infty$ one finds
$$
r_*[r] \sim - \frac{1}{\lambda} \int \frac{dr}{r^2} = \bar{r}_* +
\frac{1}{\lambda r},
$$
which leads to
\begin{eqnarray*}
V_{\mathsf{T}} (r) &\sim& \frac{d(d-2)}{4 (r_*-\bar{r}_*)^2} =\frac{\omega[(d-1)^2-1]}{4 (z-\bar{z})^2}, \\
V_{\mathsf{V}} (r) &\sim& \frac{(d-2)(d-4)}{4 (r_*-\bar{r}_*)^2} = \frac{\omega[(d-3)^2-1]}{4 (z-\bar{z})^2}, \\
V_{\mathsf{S}} (r) &\sim& \frac{(d-4)(d-6)}{4 (r_*-\bar{r}_*)^2} =
\frac{\omega[(d-5)^2-1]}{4 (z-\bar{z})^2},
\end{eqnarray*}
as one defines $z=\omega r_*$ and $\bar{z}=\omega \bar{r}_*$.

\medskip

\noindent \underline{\textsf{The RN dS Case:}} The background
space-time metric is
$$
f(r)=1-\frac{2 m}{r^{d-3}}+\frac{ q^2}{r^{2d-6}}-\lambda r^2,
$$
with $\lambda>0$. Near the black hole singularity ($r \sim 0$), the
tortoise coordinate can be expanded as
$$
r_*=\int \frac{dr}{f(r)}=\frac{1}{2d-5}\frac{r^{2d-5}}{ q^2}+\frac{2
m}{3d-8}\frac{r^{3d-8}}{ q^4}+\cdots.
$$
where we have kept the second term in the expansion of $r$ and have
chosen the integration constant so that $r_*=0$ at $r=0$, and we
have assumed $r/r_0^-\ll1$ ( $r_0^-=\left( m-\sqrt{ m^2-
q^2}\right)^{\frac1{d-3}}$ represents the inner horizon of the RN
black hole). Using this formula, one may obtain an approximate
expression of $r$ in terms of the tortoise coordinate $r_*$
\begin{equation}\label{approrrnds}
\frac r{r_0^-}\sim [(2d-5)
q^2(r_0^-)^{5-2d}r_*]^{\frac1{2d-5}}-\frac{2 m(r_0^-)^{d-3}}{(3d-8)
q^2}[(2d-5) q^2(r_0^-)^{5-2d}r_*]^{\frac{d-2}{2d-5}}.
\end{equation}
As one applies (\ref{approrrnds}) to the five different potentials
(eqs. (\ref{tensor}), (\ref{vector-charged}) and
(\ref{scalar-charged})), these potentials can be expanded near the
black hole singularity in terms of the tortoise coordinate,
respectively, as
\begin{eqnarray*}
V_{\mathsf{T}} (r) &\sim& - \frac{1}{4 z^{2}}\left[1-j_T^2-W_{RNdST}\left(\frac{z}{(r_0^-)^{2d-5}\omega}\right)^{\frac{d-3}{2d-5}}\right]\\
V_{\mathsf{V}^{\pm}} (r) &\sim& - \frac{1}{4 z^{2}}\left[1-j_{V^{\pm}}^2-W_{RNdSV^{\pm}}\left(\frac{z}{(r_0^-)^{2d-5}\omega}\right)^{\frac{d-3}{2d-5}}\right]  \\
V_{\mathsf{S}^{\pm}} (r) &\sim& - \frac{1}{4
z^{2}}\left[1-j_{S^{\pm}}^2-W_{RNdSS^{\pm}}\left(\frac{z}{(r_0^-)^{2d-5}\omega}\right)^{\frac{d-3}{2d-5}}\right],
\end{eqnarray*}
where
$$
j_T= j_{S^{\pm}}= \frac{d-3}{2d-5},\ \ \ \
j_{V^{\pm}}=\frac{3d-7}{2d-5}=2-j_T,
$$
and we have rescaled the tortoise coordinate $z = \omega r_*$. The
expression of $W_{RNdST}$, $W_{RNdSV^{\pm}}$ and $W_{RNdSS^{\pm}}$
in these equations are
\begin{eqnarray*}
W_{RNdST} &=& 0,
\\
W_{RNdSV^{\pm}} &=& \frac{4 m(d-2)(d^2-d-4)-4(3d-8)[(d-1)
m\mp\Delta]}{(2d-5)(3d-8)}(r_0^-)^{d-3}\left[(2d-5)
q^2\right]^{\frac{2-d}{2d-5}},
\\
W_{RNdSS^{\pm}} &=& -[2 m(d-2)(4d-10)
q^2+\mathcal{W}_{\pm}](r_0^-)^{d-3}\left[(2d-5)
q^2\right]^{\frac{7-3d}{2d-5}},
\end{eqnarray*}
where
\begin{eqnarray*}
\mathcal{W}_+ &=&
- m q^2[-(d-1)(3d-8)(1-\Omega)+2(4d^2-15d+12)]-2(d-2)(3d-8) m q^2-\\
&&-\frac{4(3d-8)[\ell(\ell+d-3)-(d-2)] q^4}{(d-1)(1-\Omega) m},\\
\mathcal{W}_- &=&
- m q^2[(d-1)(3d-8)(1-\Omega)+2(d-2)^2]-2(d-2)(3d-8) m q^2-\\
&&-\frac{4(3d-8)[\ell(\ell+d-3)-(d-2)] q^4}{(d-1)(1+\Omega) m}.
\end{eqnarray*}
This is just like the pure RN case.

Near $r = \infty$ one finds
$$
r_*[r] \sim - \frac{1}{\lambda} \int \frac{dr}{r^2} = \bar{r}_* +
\frac{1}{\lambda r},
$$
which leads to
\begin{eqnarray*}
V_{\mathsf{T}} (r) &\sim& \frac{d(d-2)}{4 (r_*-\bar{r}_*)^2} =\frac{\omega[(d-1)^2-1]}{4 (z-\bar{z})^2}, \\
V_{\mathsf{V}} (r) &\sim& \frac{(d-2)(d-4)}{4 (r_*-\bar{r}_*)^2} = \frac{\omega[(d-3)^2-1]}{4 (z-\bar{z})^2}, \\
V_{\mathsf{S}} (r) &\sim& \frac{(d-4)(d-6)}{4 (r_*-\bar{r}_*)^2} =
\frac{\omega[(d-5)^2-1]}{4 (z-\bar{z})^2},
\end{eqnarray*}
as one defines $z=\omega r_*$ and $\bar{z}=\omega \bar{r}_*$.

\medskip

\noindent \underline{\textsf{The Schwarzschild AdS Case:}} In this
case, the black hole is uncharged, so one has
$V_{\mathsf{V^+}}=V_{\mathsf{V^-}}$ and
$V_{\mathsf{S^+}}=V_{\mathsf{S^-}}$. For convenience, we denote them
by $V_{\mathsf{V}}$ and $V_{\mathsf{S}}$, respectively. The
background space-time metric is
$$
f(r)=1-\frac{2 m}{r^{d-3}}-\lambda r^2,
$$
with $\lambda<0$. Near the black hole singularity ($r \sim 0$), the
tortoise coordinate can be expanded as
$$
r_*=\int \frac{dr}{f(r)}=-\frac{1}{d-2}\frac{r^{d-2}}{2
m}-\frac{1}{2d-5}\frac{r^{2d-5}}{(2 m)^2}+\cdots,
$$
where we have kept the second term in the expansion of $r$ and have
chosen the integration constant so that $r_*=0$ at $r=0$. Using this
formula, one may obtain an approximate expression of $r$ in terms of
the tortoise coordinate $r_*$
\begin{equation}\label{approrsads}
r\sim[-2 m (d-2)r_*]^{\frac1{d-2}}+\frac{(d-2)r_*}{2d-5}.
\end{equation}
As one applies (\ref{approrsads}) to the three different potentials
(eqs. (\ref{tensor}), (\ref{vector-charged}) and
(\ref{scalar-charged})), these potentials can be expanded near the
black hole singularity in terms of the tortoise coordinate,
respectively, as
\begin{eqnarray*}
V_{\mathsf{T}} (r) &\sim& - \frac{1}{4 z^{2}}\left[1-0^2-W_{SAdST}\left(\frac{z}{\omega}\right)^{\frac{d-3}{d-2}}\right]\\
V_{\mathsf{V}} (r) &\sim& - \frac{1}{4 z^{2}}\left[1-2^2-W_{SAdSV}\left(\frac{z}{\omega}\right)^{\frac{d-3}{d-2}}\right]  \\
V_{\mathsf{S}} (r) &\sim& - \frac{1}{4
z^{2}}\left[1-0^2-W_{SAdSS}\left(\frac{z}{\omega}\right)^{\frac{d-3}{d-2}}\right],
\end{eqnarray*}
where
\begin{eqnarray*}
W_{SAdST} &=& \frac1{[-2(d-2)
m]^{\frac1{d-2}}}\left[\frac{2(d-3)^2}{2d-5}+\frac{4\ell(\ell+d-3)}{d-2}\right],
\\
W_{SAdSV} &=& \frac1{[-2(d-2)
m]^{\frac1{d-2}}}\left[\frac{2(d^2-8d+13)}{2d-15}+\frac{4\ell(\ell+d-3)}{d-2}\right],
\\
W_{SAdSS} &=& \frac1{[-2(d-2)
m]^{\frac1{d-2}}}\left[\frac{2d^3-24d^2+94d-116}{(2d-5)(d-2)}+\frac{4(d^2-7d+14)[\ell(\ell+d-3)-d+2]}{(d-1)(d-2)^2}\right],
\end{eqnarray*}
and we have rescaled the tortoise coordinate $z = \omega r_*$. It is
easily seen that this is just like in the pure Schwarzschild case.

Near $r = \infty$ one finds
$$
r_*[r] \sim - \frac{1}{\mid\lambda\mid} \int \frac{dr}{r^2} =
\bar{r}_* + \frac{1}{\mid\lambda\mid r},
$$
which leads to
\begin{eqnarray*}
V_{\mathsf{T}} (r) &\sim& \frac{d(d-2)}{4 (r_*-\bar{r}_*)^2} =\frac{\omega[(d-1)^2-1]}{4 (z-\bar{z})^2}, \\
V_{\mathsf{V}} (r) &\sim& \frac{(d-2)(d-4)}{4 (r_*-\bar{r}_*)^2} = \frac{\omega[(d-3)^2-1]}{4 (z-\bar{z})^2}, \\
V_{\mathsf{S}} (r) &\sim& \frac{(d-4)(d-6)}{4 (r_*-\bar{r}_*)^2} =
\frac{\omega[(d-5)^2-1]}{4 (z-\bar{z})^2},
\end{eqnarray*}
as one defines $z=\omega r_*$ and $\bar{z}=\omega \bar{r}_*$.

\medskip

\noindent \underline{\textsf{The RN AdS Case:}}  The background
space-time metric is
$$
f(r)=1-\frac{2 m}{r^{d-3}}+\frac{ q^2}{r^{2d-6}}-\lambda r^2,
$$
with $\lambda<0$. Near the black hole singularity ($r \sim 0$), the
tortoise coordinate can be expanded as
$$
r_*=\int \frac{dr}{f(r)}=\frac{1}{2d-5}\frac{r^{2d-5}}{ q^2}+\frac{2
m}{3d-8}\frac{r^{3d-8}}{ q^4}+\cdots.
$$
where we have kept the second term in the expansion of $r$ and have
chosen the integration constant so that $r_*=0$ at $r=0$, and we
have assumed $r/r_0^-\ll1$ ( $r_0^-=\left( m-\sqrt{ m^2-
q^2}\right)^{\frac1{d-3}}$ represents the inner horizon of the RN
black hole). Using this formula, one may obtain an approximate
expression of $r$ in terms of the tortoise coordinate $r_*$
\begin{equation}\label{approrrnads}
\frac r{r_0^-}\sim [(2d-5)
q^2(r_0^-)^{5-2d}r_*]^{\frac1{2d-5}}-\frac{2 m(r_0^-)^{d-3}}{(3d-8)
q^2}[(2d-5) q^2(r_0^-)^{5-2d}r_*]^{\frac{d-2}{2d-5}}.
\end{equation}
As one applies (\ref{approrrnads}) to the five different potentials
(eqs. (\ref{tensor}), (\ref{vector-charged}) and
(\ref{scalar-charged})), these potentials can be expanded near the
black hole singularity in terms of the tortoise coordinate,
respectively, as
\begin{eqnarray*}
V_{\mathsf{T}} (r) &\sim& - \frac{1}{4 z^{2}}\left[1-j_T^2-W_{RNAdST}\left(\frac{z}{(r_0^-)^{2d-5}\omega}\right)^{\frac{d-3}{2d-5}}\right]\\
V_{\mathsf{V}^{\pm}} (r) &\sim& - \frac{1}{4 z^{2}}\left[1-j_{V^{\pm}}^2-W_{RNAdSV^{\pm}}\left(\frac{z}{(r_0^-)^{2d-5}\omega}\right)^{\frac{d-3}{2d-5}}\right]  \\
V_{\mathsf{S}^{\pm}} (r) &\sim& - \frac{1}{4
z^{2}}\left[1-j_{S^{\pm}}^2-W_{RNAdSS^{\pm}}\left(\frac{z}{(r_0^-)^{2d-5}\omega}\right)^{\frac{d-3}{2d-5}}\right],
\end{eqnarray*}
where
$$
j_T= j_{S^{\pm}}= \frac{d-3}{2d-5},\ \ \ \
j_{V^{\pm}}=\frac{3d-7}{2d-5}=2-j_T,
$$
and we have rescaled the tortoise coordinate $z = \omega r_*$. The
expression of $W_{RNAdST}$, $W_{RNAdSV^{\pm}}$ and
$W_{RNAdSS^{\pm}}$ in these equations are
\begin{eqnarray*}
W_{RNAdST} &=& 0,
\\
W_{RNAdSV^{\pm}} &=& \frac{4 m(d-2)(d^2-d-4)-4(3d-8)[(d-1)
m\mp\Delta]}{(2d-5)(3d-8)}(r_0^-)^{d-3}\left[(2d-5)
q^2\right]^{\frac{2-d}{2d-5}},
\\
W_{RNAdSS^{\pm}} &=& -[2 m(d-2)(4d-10)
q^2+\mathcal{W}_{\pm}](r_0^-)^{d-3}\left[(2d-5)
q^2\right]^{\frac{7-3d}{2d-5}},
\end{eqnarray*}
where
\begin{eqnarray*}
\mathcal{W}_+ &=&
- m q^2[-(d-1)(3d-8)(1-\Omega)+2(4d^2-15d+12)]-2(d-2)(3d-8) m q^2-\\
&&-\frac{4(3d-8)[\ell(\ell+d-3)-(d-2)] q^4}{(d-1)(1-\Omega) m},\\
\mathcal{W}_- &=&
- m q^2[(d-1)(3d-8)(1-\Omega)+2(d-2)^2]-2(d-2)(3d-8) m q^2-\\
&&-\frac{4(3d-8)[\ell(\ell+d-3)-(d-2)] q^4}{(d-1)(1+\Omega) m}.
\end{eqnarray*}
This is just like the pure RN case.

Near $r = \infty$ one finds
$$
r_*[r] \sim - \frac{1}{\mid\lambda\mid} \int \frac{dr}{r^2} =
\bar{r}_* + \frac{1}{\mid\lambda\mid r},
$$
which leads to
\begin{eqnarray*}
V_{\mathsf{T}} (r) &\sim& \frac{d(d-2)}{4 (r_*-\bar{r}_*)^2} =\frac{\omega[(d-1)^2-1]}{4 (z-\bar{z})^2}, \\
V_{\mathsf{V}} (r) &\sim& \frac{(d-2)(d-4)}{4 (r_*-\bar{r}_*)^2} = \frac{\omega[(d-3)^2-1]}{4 (z-\bar{z})^2}, \\
V_{\mathsf{S}} (r) &\sim& \frac{(d-4)(d-6)}{4 (r_*-\bar{r}_*)^2} =
\frac{\omega[(d-5)^2-1]}{4 (z-\bar{z})^2},
\end{eqnarray*}
as one defines $z=\omega r_*$ and $\bar{z}=\omega \bar{r}_*$.

%%%%%%%%%%%%%%%%%%%%%%%%%%%%%%%%%%%%%%%%%%%%%%%%%%%%%%%%%%%%%%%%%
%%%%%%%%%%%%%%%%%%%%%%%%%%%%%%%%%%%%%%%%%%%%%%%%%%%%%%%%%%%%%%%%%

\vfill

\eject

\bibliographystyle{plain}
%\bibliography{papers}

\end{document}